\documentclass[showpacs,aps,pra,twocolumn,superscriptaddress]{revtex4-2}

\usepackage[english]{babel}
\usepackage{graphicx}
\usepackage{comment}
\usepackage{hyperref}
\usepackage{epsfig}
\usepackage{color}
\usepackage{xcolor} 
\usepackage{physics}
\usepackage{amssymb}
\usepackage{amsmath}
\usepackage{mathtools}
\usepackage{stmaryrd}
\usepackage{array}
\usepackage{soul}
\usepackage{flushend}
\usepackage{enumitem}
\usepackage{ytableau}
\usepackage{dsfont}
\usepackage{float}

\hypersetup{
	colorlinks,
	linkcolor={red!95!black},
	citecolor={green!60!black},
	urlcolor={blue!95!black}
}

\newcommand{\Nf}{N_{\rm f}}

\begin{document}
%
\title{$\mathrm{SU}(3)$ Fermi-Hubbard gas with three-body losses: symmetries and dark states}

%
%
\author{Alice March\'e}
\affiliation{Université Paris-Saclay, CNRS, LPTMS, 91405, Orsay, France}
\author{Alberto Nardin}
\affiliation{Université Paris-Saclay, CNRS, LPTMS, 91405, Orsay, France}
\author{Hosho Katsura}
\affiliation{Department of Physics, Graduate School of Science, The University of Tokyo, 7-3-1 Hongo, Tokyo 113-0033, Japan}
\affiliation{Institute for Physics of Intelligence, The University of Tokyo, 7-3-1 Hongo, Tokyo 113-0033, Japan}
\affiliation{Trans-scale Quantum Science Institute, The University of Tokyo, 7-3-1, Hongo, Tokyo 113-0033, Japan}
\author{Leonardo Mazza}
\affiliation{Université Paris-Saclay, CNRS, LPTMS, 91405, Orsay, France}
\affiliation{Institut Universitaire de France, 75005, Paris, France}
\date{\today}
%
\begin{abstract}
We study an $\mathrm{SU}(3)$ invariant Fermi-Hubbard gas undergoing on-site three-body losses. The model presents eight independent strong symmetries preventing the complete depletion of the gas. By making use of a basis of semi-standard Young tableaux states, we reveal the presence of a rich phenomenology of stationary states. We classify the latter according to the irreducible representation of $\mathrm{SU}(3)$ to which they belong. We finally discuss the presence of three-particle stationary states that are not protected by the $\mathrm{SU}(3)$ symmetry.
\end{abstract}
%
\maketitle 
%
%
\section{Introduction}
\label{sec:intro}
Correlated losses of particles in ultracold gases give rise to a large variety of intriguing phenomena~\cite{Fazio2024}. 
For instance, an early experiment has shown that the dynamics of the gas can be slowed down at large loss rates by the continuous quantum Zeno effect, and that when the particles are bosons, the gas should undergo fermionization~\cite{Syassen2008,Durr2009,Barontini2013, Yan2013,Zhu2014,Rossini2021}. For weak loss rates, instead, the rapidity distribution of one-dimensional Bose gases features a non-trivial decay~\cite{Bouchoule2020,Riggio2024}, and the system is driven out of equilibrium, so that Tan's relation breaks down~\cite{Bouchoule2021}. When bosons are trapped in a harmonic potential, two-body losses induce collective motions and coherent oscillations in both position and momentum spaces~\cite{Rosso2022bis,Maki2024}. Additionally, for both bosons and fermions, the decay of particle density has been found to exhibit emergent universal features~\cite{Rosso2023,Yoshida2023,Begg2024,Marche2024}.

The spin degrees of freedom play a crucial role in the depletion dynamics of fermionic gases. Indeed, losses typically occur when two particles are at the same position, which constrains their orbital wave function to be symmetric under particle exchange. Consequently, since the particles obey fermionic statistics, their spin wave function must be antisymmetric. If that is not the case, spin degrees of freedom generally prevent the complete depletion of the gas, and the dissipative dynamics give rise to spin-entangled steady states~\cite{Foss-Feig2012,Sponselee2019,Nakagawa2020,Honda2023}.

Most studies have focused on $\mathrm{SU}(2)$-invariant fermionic gases with local two-body losses, as they represent the minimal system exhibiting this phenomenon with many-body stationary states. In this case, dynamical quantities have been analytically derived, when the particles are trapped in a lattice~\cite{Rosso2021,Yoshida2023}. Moreover, the Liouvillian spectrum can be determined using Bethe Ansatz for one-dimensional lattices~\cite{Nakagawa2021}.

However, experimental extensions to an (almost) $\mathrm{SU}(3)$-symmetric gas with two-body or three-body losses have been realized by manipulating a gas of $^6\mathrm{Li}$ under strong magnetic fields~\cite{Ottenstein2008,Williams2009,Huckans2009,Wenz2009, Lompe2010,Schumacher2023} and motivate novel theoretical works~\cite{Rosso2022}. 

In the context of closed systems, the theoretical study of the $\mathrm{SU}(N)$ Fermi-Hubbard model, for $N>2$, has attracted great
interest~\cite{Honerkamp2004,honerkamp2004bcs,rapp2007color,capponi2008molecular,cazalilla2009ultracold,katsura2013nagaoka,Yoshida2021bis,Tamura2021,Kim2024}, and stimulated the development of advanced numerical techniques~\cite {Nataf2024ED,Nataf2024Nagaoka,Nataf2024DMRG}. Among the various existing results, we can cite the analysis of the metal-insulator transition~\cite{Assaraf1999, Ibarra2023, Feng2023}, the study of the large $N$ limit~\cite{marston1989large,Xu2018,Nataf2024TwoSites}, and the presence of symmetry-protected topological phases~\cite{CAPPONI201650}. Additionally, it is now well-established that such systems up to $N=10$ can be realized experimentally with ultracold alkaline-earth atoms trapped on optical lattices~\cite{Gorshkov2010,taie2010realization,taie20126,Cazalilla2014,hofrichter2016direct,Ibarra-Garcia-Padilla2025}.

In this article, we study the case of a $\mathrm{SU}(3)$ invariant fermionic gas confined on a one-dimensional lattice with local three-body losses. 
In the first part of the article we show that, in general, the $\mathrm{SU}(3)$-invariance of the model prevents the system from becoming completely empty at late time. We write a basis of the Hilbert space in the form of semi-standard Young tableaux which takes care of the symmetries of the model. We determine how the Young tableaux states are modified by three-body loss processes, enabling us to derive a lower bound for the average number of fermions remaining on the lattice, which holds at all times during the dynamics. 

In the second part of the article, we discuss the stationary states of the model. We show that, in this $\mathrm{SU}(3)$-symmetric case with three-body losses, the zoology of stationary states is much richer than for a $\mathrm{SU}(3)$ or $\mathrm{SU}(2)$-symmetric Fermi-Hubbard Hamiltonian with two-body losses~\cite{Foss-Feig2012, Rosso2022}. In the latter case, the stationary states are necessarily factorized into orbital and spin parts in first quantization. With only three-body losses, this constraint is lifted. We classify the stationary states according to the irreducible representation of $\mathrm{SU}(3)$ they belong to. We provide explicit expressions for some of them. Finally, we discuss the presence of dark states, for periodic boundary conditions, in sectors of the Hilbert space where losses are not \textit{a priori} prevented by symmetries. Remarkably, these non-trivial dark states emerge due to quantum interference effects.

The article is organized as follows. In Sec.~\ref{sec:model}, we describe the model. In Sec.~\ref{sec:symmetries}, we show that the Hamiltonian is invariant under any global $\mathrm{SU}(3)$ rotation in spin space, and that the expectation values of the eight $\mathrm{SU}(3)$ pseudo-spin algebra generators are conserved during the lossy dynamics. Then, we review some known results about the symmetry-relevant basis of states which take the form of semi-standard Young tableaux. In Sec.~\ref{sec:Effect_of_diss}, we discuss new results concerning the effect of a loss on the Young tableau states, unveiling some other quantities conserved during a loss process. As a direct consequence, a lower bound for the average number of fermions is derived. In Sec.~\ref{sec:DS}, we give expressions of dark states belonging to each irreducible representations of $\mathrm{SU}(3)$, and we show how to count them. In Sec.~\ref{sec:NANBNC1}, we analyze in detail the case of three fermions of three different spin components.
The general validity of our result in higher-dimensional lattices or in the presence of a confining potential is discussed in Sec.~\ref{Sec:Discussion}.
The conclusions of our results are presented in Sec.~\ref{sec:conclu}. Some technical details are deferred to eleven Appendices. 

\section{Model} 
\label{sec:model}

We consider a one-dimensional lattice of $L$ sites populated by three-spin-component fermions that can hop between neighboring lattice sites and feature contact interaction.
We denote $c_{j \sigma}^{\dag}$ and $c_{j \sigma}$ the creation and annihilation operators of a fermion at site $j \in \{ 1 , \ldots ,L \}$ with spin $\sigma \in \{ A,B,C \}$, satisfying the canonical anticommutation relations
\begin{subequations}
\label{eq:anticomm}
\begin{align}
    \{ c_{j\sigma}^\dag,c_{j\sigma^\prime}^\dag \}&= \{ c_{j\sigma},c_{i\sigma^\prime} \}=0,\\
    \{ c_{j\sigma}^\dag,c_{i\sigma^\prime} \}&= \delta_{ij} \delta_{\sigma \sigma^\prime}.
\end{align}
\end{subequations}
This system is described by the Fermi-Hubbard Hamiltonian
\begin{align}
H_{\rm Hub} = H_{\rm hop} + H_{\rm int,2}+ H_{\rm int,3}
\label{eq:H_Hub}
\end{align}
with
\begin{subequations}
\begin{align}
     H_{\rm hop} &= -J \sum_{j} \sum_{\sigma = A,B,C} \left( c_{j+1 \sigma}^\dag c_{j \sigma} + \mathrm{h.c.} \right), \label{eq:Hhop} \\
     H_{\rm int,2} &= U_2  \sum_{j} \left( n_{j A} n_{j B} + n_{j A} n_{j C} + n_{j B} n_{j C} \right), \label{eq:Hint2} \\ 
     H_{\rm int,3} &= U_3  \sum_{j} n_{j A} n_{j B}  n_{j C}. \label{eq:Hint3}
\end{align}
\end{subequations}
Here, $J$ is the hopping amplitude, $U_2$ and $U_3$ are the spin-independent two-body and three-body contact interaction strengths and $n_{j \sigma} = c_{j \sigma}^{\dag} c_{j \sigma}$ is the local spin-density operator. The three-body interaction Hamiltonian $H_{\rm int,3}$ can eventually appear in cold-atom experiments; a similar Hamiltonian is engineered in Ref.~\cite{Goban2018}. However, three-body interactions can also be absent, but since the corresponding term is $\mathrm{SU}(3)$-symmetric it can be included without any difficulty in our theoretical study which concerns only the stationary properties of the gas and not its dynamics. The reader may consider $U_3 = 0$ or $U_3 \neq 0$ at will, without affecting either the $\mathrm{SU}(3)$-symmetry or the stationary states of the system (since they cannot contain any triple lattice site occupancy, as it will be rigorously shown in Sec.~\ref{sec:DSandTheorems}).

As customary, we model the time evolution of the density matrix $\rho$ of the gas in the lattice according to a Lindblad master equation
\begin{equation}
    \frac{d \rho}{ dt} = - i \left[ H_{\rm Hub} , \rho\right] + \sum_{j} \left( L_j \rho L_j^\dag - \frac{1}{2} \{ L_j^\dag L_j , \rho \} \right) =: \mathcal{L}[\rho]. \label{eq:MElattice}
\end{equation}
Here, $\mathcal{L}$ is the Liouvillian superoperator acting on $\rho$.
We assume that the previously described fermionic gas is subject to local three-body losses modeled by the jump operators 
\begin{equation}
    L_{j} = \sqrt{\gamma} c_{jA} c_{jB} c_{jC} \label{eq:Lj},
\end{equation} where $\gamma$ is the loss rate. 
Throughout this article, we set~$\hbar = 1$.

We remark that fermionic gases with three-spin components can be realized experimentally via ultracold alkaline-earth atoms such as $^{173}\mathrm{Yb}$ or $^{87}\mathrm{Sr}$~\cite{Gorshkov2010,Cazalilla2014,Ibarra-Garcia-Padilla2025}. Indeed, in the ground state, the total electronic angular momentum of these atoms $J$ vanishes. Thus, the nuclear spin $I$ perfectly decouples from the electronic structure, and the total angular momentum of the atom is $F=I$. Since the atoms interact only via their electronic shells and the spin degrees of freedom are inside the nucleus, the interaction is $\mathrm{SU}(N)$-symmetric, with $N$ the number of possible spin components. Experimentally, $N$ can be tuned up to $N=2F+1$, by selectively populating the states $\ket{m_F = -F},\ldots, \ket{m_F = F}$. To date, we are not aware of any experimental result on three-body losses in such $\mathrm{SU}(3)$ gases  made of alkaline-earth atoms.

Another experimental technique that could be used to implement our $\mathrm{SU}(3)$-invariant setup involves the manipulation of ultracold $^{6}\mathrm{Li}$ atoms~\cite{Ottenstein2008,Williams2009,Huckans2009,Wenz2009, Lompe2010,Schumacher2023}. When an $^{6}\mathrm{Li}$ atom is subject to an increasingly strong magnetic field, its three lowest Zeeman hyperfine levels ($\ket{F=1/2, m_F = 1/2}$, $\ket{F=1/2, m_F = -1/2}$ and $\ket{F=3/2, m_F = -3/2}$) become better and better separated in energy from the rest of the spectrum, enabling their selective population. Furthermore, in this large bias field regime the single valence electron of the atom becomes increasingly spin-polarized, leading to an almost perfect decoupling between electronic and nuclear spins, and thus to $\mathrm{SU}(3)$-symmetric two-body elastic collisions between different atoms, with still important three-body losses~\cite{Williams2009, Schumacher2023}.
In this setup, one- and two-body losses are typically negligible compared to three-body losses. This was verified experimentally by comparing the decay rate of two-spin-component mixtures with that of an unpolarized gas containing all three spin components (see Fig.~3 in Ref.~\cite{Huckans2009} or Fig.~S2 of Ref.~\cite{Schumacher2023}).
We also mention that methods for realizing $\mathrm{SU}(3)$ invariant two-body interactions at finite bias fields have been studied~\cite{OHara_2011}.

\section{Symmetries} \label{sec:symmetries}
\subsection{Invariance of the Hamiltonian under any global $\mathrm{SU}(3)$ rotation in spin space} \label{sec:SU3rot}
In this section, we aim to show that the Hamiltonian $H_{\rm Hub}$ is invariant under any global $\mathrm{SU}(3)$ rotation in spin space. For this, we introduce the spin-independent operators~\cite{Paldus1981,Botzung2024,Botzung2024bis}
\begin{equation}
    E_{ij} = \sum_{\sigma} c_{i \sigma}^\dag c_{j \sigma} \quad \text{with } i,j \in \{1,\ldots,L\}, \label{eq:Eij}
\end{equation}
which satisfy the commutation relations of the Lie algebra of the $\mathrm{U}(L)$ generators:
\begin{equation}
    [ E_{ij},E_{kl}] = \delta_{jk} E_{il} - \delta_{il} E_{kj}.
\end{equation}
The terms of the Hamiltonian $H_{\rm Hub}$ in Eq.~\eqref{eq:H_Hub} can be rewritten with the operators~\eqref{eq:Eij} as
\begin{subequations}
\begin{align}
     H_{\rm hop} &= -J \sum_{j} \left( E_{j j+1} + \mathrm{h.c.} \right),  \\
     H_{\rm int,2} &= \frac{U_2}{2}  \sum_{j}  E_{jj} \left( E_{jj} - 1 \right),  \\ 
     H_{\rm int,3} &= \frac{U_3}{6}  \sum_{j} E_{jj} \left( E_{jj} - 1 \right) \left( E_{jj} - 2 \right) . 
\end{align} \label{eq:Hhub2}
\end{subequations}

On the other hand, we define the spin ladder operators
\begin{equation}
    F_{\alpha \beta} = \sum_j c_{j \alpha}^\dag c_{j \beta}  \quad \text{with } \alpha,\beta \in \{A,B,C\} \label{eq:Fab},
\end{equation}
which satisfy similar commutation relations:
\begin{equation}
    [ F_{\alpha \beta},F_{\gamma \sigma}] = \delta_{\beta \gamma} F_{\alpha \sigma} - \delta_{\alpha \sigma} F_{\gamma \beta}. 
\end{equation}
Importantly, for all $i,j \in \{1, \ldots, L\}$ and $\alpha,\beta \in \{ A, B, C \}$,
\begin{equation}
    [E_{ij},F_{\alpha \beta}]=0. \label{eq:EF}
\end{equation}
Eq.~\eqref{eq:EF} can be shown via the fermionic anti-commutation relations of Eq.~\eqref{eq:anticomm}. Since any global rotation in spin space $R$ is generated by the operators~\eqref{eq:Fab}, Eqs.~\eqref{eq:Hhub2} and~\eqref{eq:EF} imply that $H_{\rm Hub}$ is indeed invariant under such a rotation, i.e., $[H_{\rm Hub,}R]=0$.
\subsection{Strong symmetries}
We now identify the strong symmetries of the dissipative dynamics. By definition, a \textit{strong symmetry} $O$ is an operator which commutes with both the Hamiltonian and all the jumps operators: $[H_{\rm Hub},O]=0$ and $[L_j,O]=0$ for all $j$~\cite{Buca2012,Albert2014,Fazio2024}. It can be shown from~Eq.~\eqref{eq:MElattice} that the expectation value of a strong symmetry is conserved during the dynamics, i.e. $\frac{d}{dt} \langle O \rangle_t := \frac{d}{dt} \Tr[O \rho(t)]=0$.

As a consequence of the $\mathrm{SU}(3)$  invariance, the model has the following eight strong symmetries (see the proof in Appendix~\ref{Ap:StrongSym}):
\begin{equation}
    \Lambda_l = \frac{1}{2}\sum_j 
\begin{pmatrix}
c_{j A}^\dag & c_{j B}^\dag & c_{j C}^\dag 
\end{pmatrix}
\lambda_l 
\begin{pmatrix}
c_{j A} \\
c_{j B} \\
c_{j C}
\end{pmatrix}.
\end{equation}
Here, $\lambda_l$ ($l=1, \ldots, 8$) are the Gell-Mann matrices which span the Lie algebra of $\mathrm{SU}(3)$ and read:
\begin{align*}
\lambda_1 &=  \begin{pmatrix}
0 & 1 & 0\\
1 & 0 & 0\\
0 & 0 & 0
\end{pmatrix},\,
\lambda_2 =  \begin{pmatrix}
0 & -i & 0\\
i & 0 & 0\\
0 & 0 & 0
\end{pmatrix},\,
\lambda_3 =  \begin{pmatrix}
1 & 0 & 0\\
0 & -1 & 0\\
0 & 0 & 0
\end{pmatrix}, \\
\lambda_4 &=  \begin{pmatrix}
0 & 0 & 1\\
0 & 0 & 0\\
1 & 0 & 0
\end{pmatrix},\,
\lambda_5 =  \begin{pmatrix}
0 & 0 & -i\\
0 & 0 & 0\\
i & 0 & 0
\end{pmatrix},\,
\lambda_6 =  \begin{pmatrix}
0 & 0 & 0\\
0 & 0 & 1\\
0 & 1 & 0
\end{pmatrix}, \\ 
\lambda_7 &=  \begin{pmatrix}
0 & 0 & 0\\
0 & 0 & -i\\
0 & i & 0
\end{pmatrix},\, 
\lambda_8 =  \frac{1}{\sqrt{3}}\begin{pmatrix}
1 & 0 & 0\\
0 & 1 & 0\\
0 & 0 & -2
\end{pmatrix}.
\end{align*}
We denote the operators $\Lambda_l$ as the pseudo-spin generators of the $\rm{SU}(3)$ group. 

By definition, an operator is called a \textit{Casimir} when it commutes with all the generators of a Lie group. In the context of $\mathrm{SU}(3)$, there are two independent and commuting Casimir operators, one quadratic and one cubic in the $\Lambda_l$~\cite{Pais1966}:
\begin{equation}
C_2 = \sum_l \Lambda_l^2 \quad \text{and} \quad C_3 = \sum_{j, k, l} d_{j k l} \Lambda_j \Lambda_k \Lambda_l,
\end{equation}
with the coefficients $d_{jkl} = \frac{1}{4} \Tr[\lambda_j \{\lambda_k, \lambda_l \}]$. Since the latter are also strong symmetries, the full Hilbert space can be decomposed into different symmetry sectors labeled by the eigenvalues of $C_2$ and $C_3$, and the expectation values of $C_2$ and $C_3$ are constants of motion.
An alternative way of labeling the symmetry sectors is done with the irreducible representations (irrep) of $\mathrm{SU}(3)$. Each irrep of $\mathrm{SU}(3)$ is itself labeled by two positive integers $(p,q)$~\cite{Georgi2000}. A state belonging to the $(p,q)$-irrep is an eigenvector of $C_2$ with eigenvalue
\begin{equation}
    c_2(p,q) = (p^2 + q^2 + 3 p + 3 q + p q)/3 \label{eq:c1}
\end{equation}
and an eigenvector of $C_3$ with eigenvalue
\begin{equation}
    c_3(p,q) = (p-q)(3 + p + 2 q)(3 + q + 2 p)/18. \label{eq:c2}
\end{equation}
This provides the correspondence between the two ways of labeling the symmetry sectors.

We also note that $\lambda_1$, $\lambda_2$, and $\lambda_3$ span a $\mathrm{SU}(2)$ subalgebra. Thus, the operators
\begin{align}
    \Lambda_1 &= \frac{1}{2} \sum_j \left( c_{j A}^\dag c_{j B}+ c_{jB}^\dag c_{jA} \right), \\
    \Lambda_2 &= \frac{i}{2} \sum_j \left( c_{j B}^\dag c_{j A} - c_{jA}^\dag c_{jB} \right), \\
    \Lambda_3 &=\frac{1}{2} \sum_j \left( n_{j A}-n_{jB} \right)
\end{align}
satisfy the commutation relations of the $\mathrm{SU}(2)$ generators: 
\begin{equation}
    [\Lambda_i,\Lambda_j]=i \epsilon_{ijk} \Lambda_k,
\end{equation}
where $\epsilon_{ijk}$ is the Levi-Civita symbol. In this $\mathrm{SU}(2)$ subalgebra, the associated total spin is
\begin{equation}
    I^2 = \Lambda_1^2 + \Lambda_2^2 + \Lambda_3^2.
\end{equation}
The latter commutes with the two Casimir operators of $\mathrm{SU}(3)$: $[I^2,C_2]=[I^2,C_3]=0$.
\subsection{Young tableaux} \label{sec:YT}
We now review some known results~\cite{Paldus1981,Nataf2024ED,Nataf2024Nagaoka} concerning the construction of a convenient orthonormal basis of the Hilbert space taking into account the symmetries of our model. 

\begin{figure}
\includegraphics[width=0.9\linewidth]{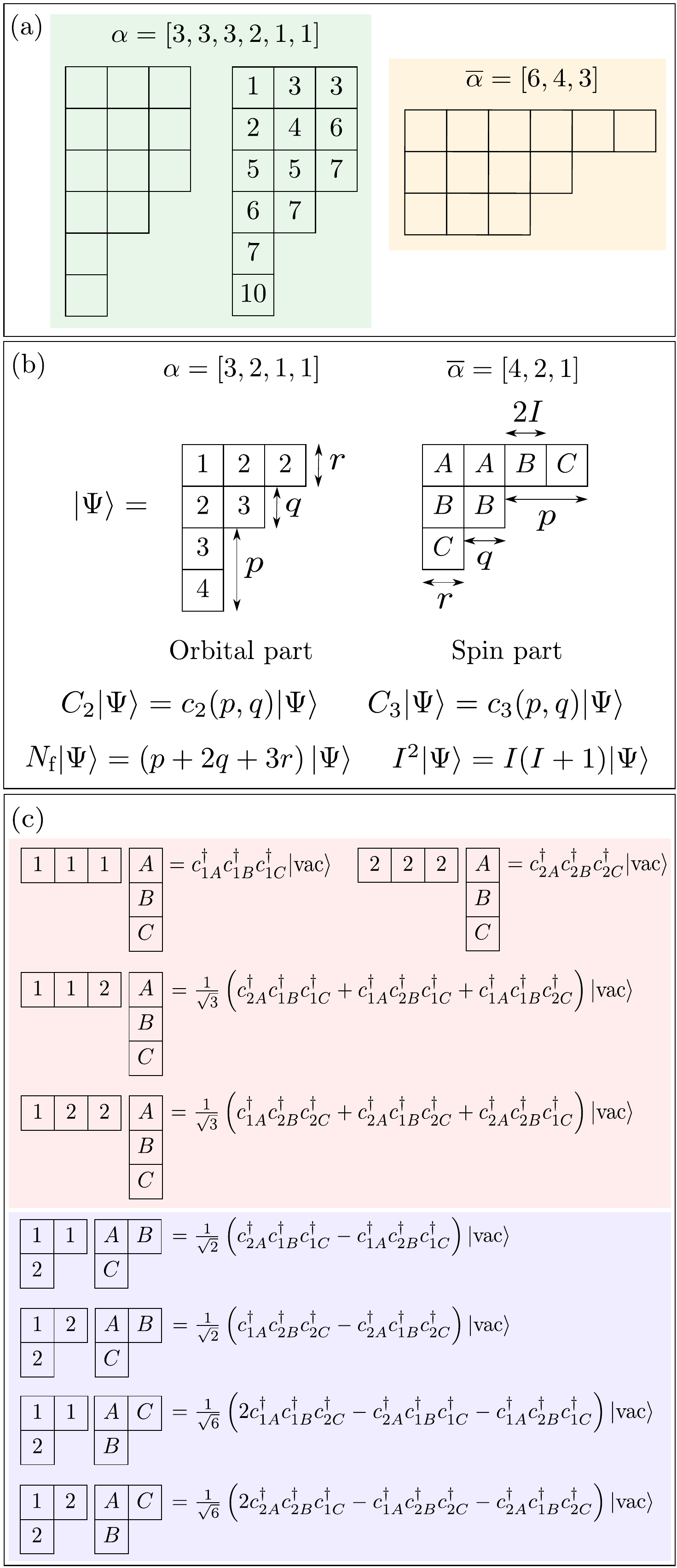}
\caption{(a) From left to right: a Young diagram of shape $\alpha = [3,3,3,2,1,1]$, a semi-standard Young tableau of the same shape, a Young diagram of shape $\overline{\alpha}=[6,4,3]$. (b) Example of a state written in terms of semi-standard Young tableaux for $\Nf = 7$ particles on $L=4$ sites, with $N_A = 2$, $N_B=3$ and $N_C=2$. From the spin part, we see that $p=2$, $q=1$, $r=1$ and $I = 1/2$. The total number of boxes in one tableau is $\Nf = p+2q+3r$. (c) Basis of the Hilbert subspace for $\Nf=3$ particles on $L=2$ lattice sites in the sector with one fermion $A$, one $B$, and one $C$. In this $8$-dimensional subspace, $4$ basis states belong to the $(p=0,q=0)$-irrep of $\mathrm{SU}(3)$ (in red) and $4$ basis states belong to the $(p=1,q=1)$-irrep (in blue).} \label{fig:example_state}
\end{figure}

For the construction of this targeted basis, Young diagrams play a central role. A \textit{Young diagram} (YD) is an ensemble of $m$ left-justified rows, with the $k$-th row consisting of $\alpha_k$ boxes. The rows are ordered in such a way that $\alpha_1 \geq \alpha_2 \geq \ldots \geq \alpha_m \geq 1$. It is customary to denote the shape of the YD as $\alpha = [\alpha_1, \alpha_2, \ldots, \alpha_m]$.
We give examples of YDs in Fig.~\ref{fig:example_state}(a). The total number of boxes $\Nf = \sum_{i=1}^m \alpha_i$ will correspond to the number of fermions in a given state. Importantly, the YDs with at most $n -1$ rows uniquely label irreps of $\mathrm{U}(n)$ or $\mathrm{SU}(n)$. A \textit{semi-standard Young tableau} (ssYT) is a YD where the boxes are filled with integers satisfying certain constraints: the entries cannot strictly decrease from left to right in each row and should strictly increase from top to bottom in each column. We give an example of ssYT in Fig.~\ref{fig:example_state}(a).

Young diagrams can very conveniently be used to describe a basis of the system's Hilbert space: each state can indeed be labeled by two ssYTs. One tableau is interpreted as the orbital part of the state and is filled with numbers from $1$ to $L$ (which in our case label the lattice sites), while the other tableau is interpreted as the spin part and is filled with the flavors $A$, $B$ and $C$. We introduce the order $A < B < C$. By construction, the tableau corresponding to the orbital part of the state belongs to an irrep of $\mathrm{U}(L)$ while the tableau corresponding to the spin part belongs to an irrep of $\mathrm{SU}(3)$.
In order to properly label a physically 
admissible state, the respective shapes of the orbital and the spin tableaux cannot be arbitrary, they have to be transposed with respect to each other.
Here, the transposition operation on a YD corresponds to a reflection along its main diagonal, thereby exchanging the roles of the columns and the rows. The latter constraint is a consequence of the fermionic nature of the gas: the state should be antisymmetric by any exchange of two particles.
If the orbital ssYT has the shape $\alpha$, we denote by $\overline{\alpha}$ the shape of the associated spin ssYT which is the transpose of $\alpha$. We give an example of a state in Fig.~\ref{fig:example_state}(b); the procedure to obtain the second quantization form of a state from its ssYT expression is in general rather non-trivial and will be explained in Sec.~\ref{sec:secondQ}.

We introduce new operators: the spin-resolved particle number $N_\sigma := \sum_j n_{j \sigma}$ ($\sigma=A$, $B$, $C$) and the total particle number $\Nf=N_A+N_B+N_C$, in terms of which we have $\Lambda_3 = (N_A -N_B)/2$ and $\Lambda_8 = (N_A + N_B -2 N_C)/2\sqrt{3}$. Each ssYT state is a simultaneous eigenstate of the mutually commuting operators~$C_2$, $C_3$, $I^2$, $\Lambda_3$, $\Lambda_8$ and~$\Nf$. As shown in Eqs.~\eqref{eq:c1} and \eqref{eq:c2}, the eigenvalues of $C_2$ and $C_3$ are linked to the integers $p$ and $q$; while the eigenvalue of~$\Nf$ is the total number of boxes in one of the two tableaux. In fact, $p$ is the number of columns with one box, while $q$ is the number of columns with two boxes in the ssYT for the spin~\cite{Georgi2000}, as shown in Fig.~\ref{fig:example_state}(b). Thus, for a given state, the eigenvalues of $C_2$, $C_3$ and $\Nf$ fully characterize the shape of the ssYTs; and we have $\Nf = p+ 2q +3r$ where $r$ is the number of columns with three boxes in the spin ssYT.
 
From a given spin ssYT, we can also directly read the corresponding eigenvalue of~$I^2$. First, we construct a new ssYT by erasing the boxes containing a $C$ in the original one. Therefore, we obtain a state belonging to an irrep of the $\mathrm{SU}(2)$ subalgebra for $A$ and $B$. From the known results on the $\mathrm{SU}(2)$ Lie algebra~\cite{Paldus1981}, we deduce that the eigenvalue of $I^2$ is written as $I(I+1)$ where $2 I$ is the number of columns with one box in the new tableau; this is illustrated in Fig.~\ref{fig:example_state}(b). $\Lambda_3$ and $\Lambda_8$ give additional information about the imbalance between the three flavors. Thus, once $(p,q,r)$ are fixed, the filling of the tableau for the spin is fully characterized by $\Lambda_3$, $\Lambda_8$ and $I^2$. 

A basis of the Hilbert subspace for $\Nf$ fermions on $L$ lattice sites is composed of the ensemble of ssYTs pairs with $\Nf$ boxes such that the orbital ssYT belongs to~$\mathrm{U}(L)$. In Fig.~\ref{fig:example_state}(c), we give an example of basis for $L=2$ and $\Nf=3$ in the subspace where there is one $A$, one $B$ and one $C$ fermions, which is composed of eight states.

\section{Effect of the dissipation} \label{sec:Effect_of_diss}
In this section, we characterize the effect of a loss process on a basis state written in terms of ssYTs. To do that, we shall obtain a writing of such a state in second quantization, i.e., as a linear combination of Slater determinants. Then, we deduce a lower bound for the total number of fermions remaining in the lattice.

Sections~\ref{sec:secondQ} and~\ref{sec:conservedQuantLoss} contain the technical details that are necessary to prove the results listed in Sec.~\ref{sec:effect_loss_YT}. The reader uninterested in these technicalities can directly go to Sec.~\ref{sec:effect_loss_YT} without affecting the understanding of the rest.

\subsection{Second quantization writing of the ssYT states} \label{sec:secondQ}

\begin{figure}[t]
\includegraphics[width=\linewidth]{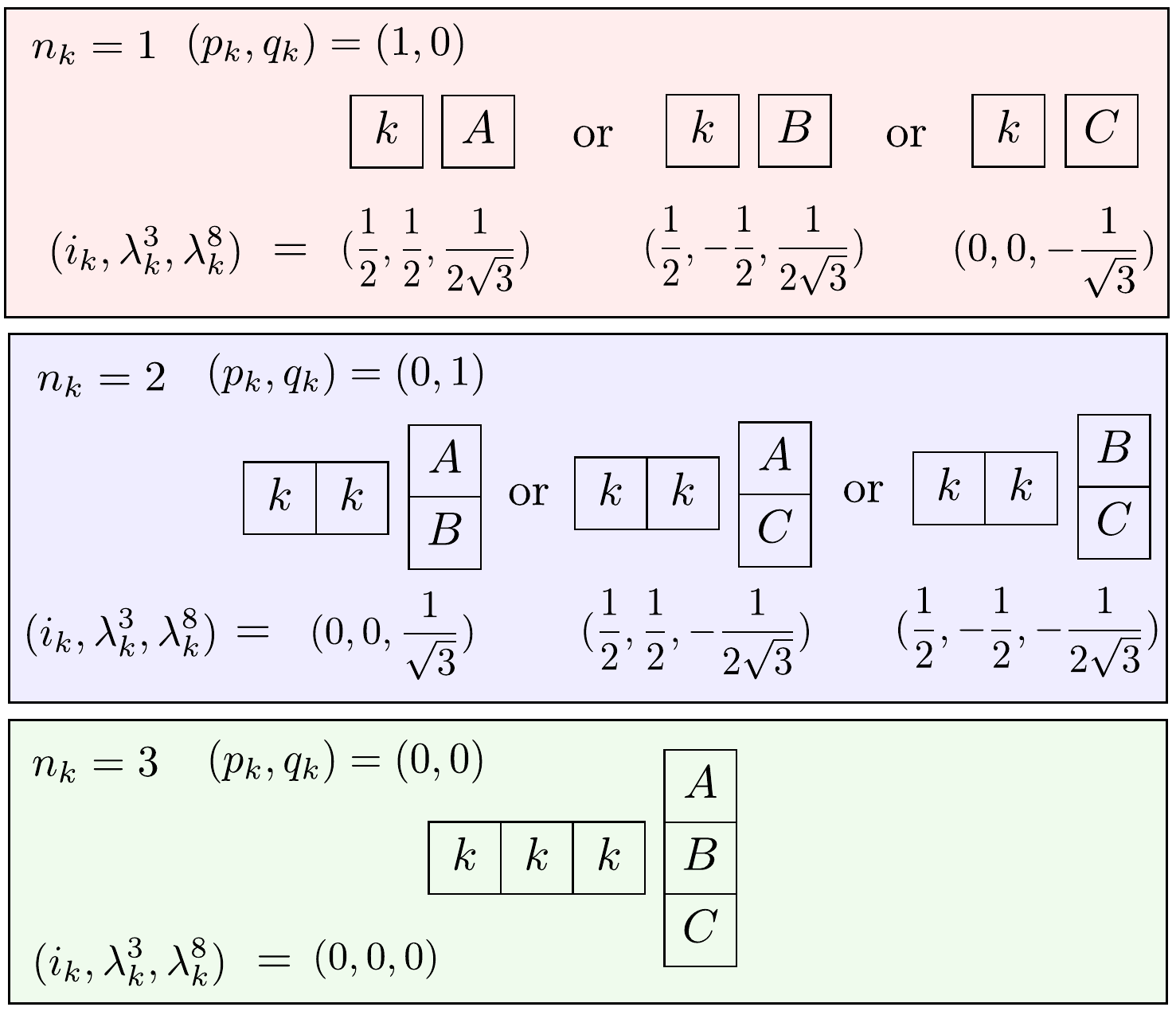}
\caption{Possible states added at step $k$ corresponding to $n_k$ fermions located at the lattice site $k$ and belonging to the $(p_k,q_k)$-irrep of $\mathrm{SU}(3)$. Here, $i_k(i_k+1)$, $\lambda_k^3$ and $\lambda_8^k$ are the eigenvalues of the operators $I^2$, $\Lambda_3$ and $\Lambda_8$ respectively.} \label{fig:state_cluster}
\end{figure}

We derive a second quantization writing of the ssYT states, by adapting some results for $\rm{SU}(2)$ spins~\cite{Moshinsky1971,Paldus1981}.
First, we construct the generalized Paldus (or ABCD) tableau associated to a given orbital ssYT. As explained in Sec.~\ref{sec:YT}, the latter has at most $L$ rows and $3$ columns, and belongs to an irrep of $U(L)$; we denote it~$\mathcal{T}_L$. In this tableau, we call $a_L$, $b_L$, $c_L$, and $d_L$ the numbers of rows with $3$, $2$, $1$, and $0$ boxes, respectively, such that $a_L+b_L+c_L+d_L = L$. Then, we construct a second ssYT, denoted~$\mathcal{T}_{L-1}$, by removing the boxes containing the integer $L$ in $\mathcal{T}_{L}$. This provides a tableau belonging to an irrep of $U(L-1)$. We call $a_{L-1}$, $b_{L-1}$, $c_{L-1}$, and $d_{L-1}$ the number of rows with $3$, $2$, $1$, and $0$ boxes, respectively, in $\mathcal{T}_{L-1}$ such that $a_{L-1}+b_{L-1}+c_{L-1}+d_{L-1} = L-1$. Then, we remove the boxes containing the integer $L-1$ in $\mathcal{T}_{L-1}$ to construct $\mathcal{T}_{L-2}$, etc. By repeating this same operation $L$ times, we construct the Paldus tableau:

\begin{table}[h!]
\begingroup
\renewcommand{\arraystretch}{1.13}
\begin{tabular}{c|c|c|c}
 $A$  & $B$ & $C$ & $D$ \\
 \hline
 \hline
 $a_L$ & $b_L$ & $c_L$ & $d_L$ \\
  \hline
 $a_{L-1}$ & $b_{L-1}$ & $c_{L-1}$ & $d_{L-1}$ \\
  \hline
 $\vdots$ & $\vdots$ & $\vdots$ & $\vdots$ \\
  \hline
$a_1$ & $b_1$ & $c_1$ & $d_1$ \\
\end{tabular}
\endgroup
\end{table}

By construction, we have, $\forall k \in \{ 1, \ldots, L \}$,
\begin{align}
    a_k + b_k + c_k + d_k = k,\\
    3 a_k + 2 b_k + c_k = \sum_{j=1}^k n_j,
\end{align}
with $n_j$ the number of boxes containing $j$ in $\mathcal{T}_L$, i.e., the number of particles at site $j$. We give an example of ABCD tableau construction in Appendix~\ref{Ap:ex_ABCD}.
Since the ssYTs for the spin and the orbital must have shapes that are the transpose of each other, the irrep of $\mathrm{SU}(3)$ associated with the state is changing as we sequentially remove boxes in the orbital ssYT.
In fact, a state having $\mathcal{T}_k$ as orbital part belongs to the ($P_k$, $Q_k$)-irrep of $\mathrm{SU}(3)$ with
\begin{equation}
    c_k = P_k, \qquad b_k = Q_k.
\end{equation}
Therefore, $(P_k,Q_k)$ is the irrep of $\mathrm{SU}(3)$ 
corresponding only to the fermions on the first $k$ sites.

We can understand the construction of a given ssYT state as the successive additions of the particles belonging to a given lattice site, from site $j=1$ to site $j=L$. At each step $k \in \{ 1, \ldots, L\}$, we start from a cluster of particles belonging to the $(P_{k-1}, Q_{k-1})$-irrep of $\mathrm{SU}(3)$, and we add $0$, $1$, $2$ or $3$ particles that are at site $k$ to construct a new cluster of particles belonging to the $(P_k, Q_k)$-irrep. Consistently with previous notations, the number of particles added at step $k$ is denoted~$n_k$. However, we denote $(p_k,q_k)$ the irrep of $\mathrm{SU}(3)$ characterizing the $n_k$ particles added. Importantly, $(p_k,q_k)$ is determined by $n_k$ only:
\begin{align}
    &\text{if } n_k=0 \text{ or } 3 \quad \text{then } p_k=0 \text{ and } q_k=0, \label{eq:nk0or3}\\
    &\text{if } n_k=1 \quad \text{then } p_k=1 \text{ and } q_k=0,\\
    &\text{if } n_k=2 \quad \text{then } p_k=0 \text{ and } q_k=1.
\end{align}
Indeed, all the particles added at step $k$ are on the same site $k$, and this imposes strong constraints on their possible ssYTs; as shown in Fig.~\ref{fig:state_cluster}.
For every possible state added at step $k$, the eigenvalues of $I^2$, $\Lambda_3$ and $\Lambda_8$, denoted $i_k(i_k+1)$, $\lambda_{k}^3$ and $\lambda_{8}^3$ respectively, are also given in Fig.~\ref{fig:state_cluster}.

To find the second quantization form of a given state written as a pair of ssYTs, we use the following procedure. From the filling of the spin ssYT, we identify the eigenvalues of $I^2$, $\Lambda_3$ and $\Lambda_8$, corresponding to the full state, denoted $I_L(I_L+1)$, $\Lambda^3_L$  and $\Lambda^8_L$ respectively. For all~$k \in \{ 1 ,\ldots, L \}$ , we also identify the occupation numbers $n_k$, and the intermediate irreps of~$\mathrm{SU}(3)$~$(P_k,Q_k) = (c_k,b_k)$ from the ABCD tableau. We deduce~$(p_k,q_k)$ from the occupation numbers. Knowing all these quantum numbers, the state can be rewritten as
\begin{widetext}
\begin{equation}
    \sum_{\substack{\{ I_k, \Lambda_k^3,  \Lambda_k^8 \} \\
        \text{compatible with} \\ 
        \{ P_k, Q_k \}}}
        \quad
    \sum_{\substack{
        \{ i_k, \lambda_{k}^3,  \lambda_{k}^8 \} \\
        \text{compatible with} \\ 
        \{ p_{k}, q_{k} \}}} \quad \prod_{k=1}^L \langle  P_{k-1} \ Q_{k-1} \ I_{k-1} \ \Lambda_{k-1}^3 \ \Lambda_{k-1}^8, \ p_k  \ q_k  \ i_k \ \lambda_k^8  \ \lambda_k^3  |  P_k \ Q_k \ I_k \ \Lambda_k^3 \ \Lambda_k^8 \rangle \  c_k^\dag (n_k,i_k,\lambda_k^3,\lambda_k^8 ) \ \ket{\rm vac},\label{eq:SDwriting}
\end{equation}
where we introduced a short-hand notation for the nested sums $$\sum_{\substack{\{ I_k, \Lambda_k^3,  \Lambda_k^8 \} \\
        \text{compatible with} \\ 
        \{ P_k, Q_k \}}} = \sum_{\substack{
        \left( I_1, \Lambda_1^3,  \Lambda_1^8 \right) \\
        \text{compatible with} \\ 
        \left( P_1, Q_1 \right)}}
    \
     \ldots  \sum_{\substack{
        \left( I_{L-1}, \Lambda_{L-1}^3,  \Lambda_{L-1}^8 \right) \\
        \text{compatible with} \\ 
        \left( P_{L-1}, Q_{L-1} \right)}} \text{ and } \sum_{\substack{
        \{ i_k, \lambda_{k}^3,  \lambda_{k}^8 \} \\
        \text{compatible with} \\ 
        \{ p_{k}, q_{k} \}}} = \sum_{\substack{
        \left( i_1, \lambda_{1}^3,  \lambda_{1}^8 \right) \\
        \text{compatible with} \\ 
        \left( p_{1}, q_{1} \right)}}
    \ldots
    \sum_{\substack{
        \left( i_L, \lambda_{L}^3,  \lambda_{L}^8 \right) \\
        \text{compatible with} \\ 
        \left( p_{L}, q_{L} \right)}}.$$ 
\end{widetext}
In Eq.~\eqref{eq:SDwriting}, the $\langle P_{k-1} \ldots | \ldots  \Lambda_k^8 \rangle$ are $\mathrm{SU}(3)$ Clebsch-Gordan coefficients, $\ket{\rm vac}$ is the vacuum state, and the operators $c_k^\dag (n_k,i_k,\lambda_k^3,\lambda_k^8 )$ are defined as:
\begin{align*}
   &\mathds{1} \text{ if } n_k=0,\\
   &c_{kA}^\dag \text{ if } n_k=1 \text{ and } ( i_k, \lambda_k^3 , \lambda_k^8) = ( \frac{1}{2},\frac{1}{2},\frac{1}{2\sqrt{3}}),  \\
   &c_{kB}^\dag \text{ if } n_k=1 \text{ and } ( i_k, \lambda_k^3 , \lambda_k^8) = ( \frac{1}{2},-\frac{1}{2},\frac{1}{2\sqrt{3}}), \\
    &c_{kC}^\dag \text{ if } n_k=1 \text{ and } ( i_k, \lambda_k^3 , \lambda_k^8) = (0,0,-\frac{1}{\sqrt{3}}), \\
    &c_{kA}^\dag c_{kB}^\dag \text{ if } n_k=2 \text{ and } ( i_k, \lambda_k^3 , \lambda_k^8) = (0,0,\frac{1}{\sqrt{3}}), \\
    &c_{kA}^\dag c_{kC}^\dag \text{ if } n_k=2 \text{ and } ( i_k, \lambda_k^3 , \lambda_k^8) = (\frac{1}{2},\frac{1}{2},-\frac{1}{2\sqrt{3}}), \\
    &c_{kB}^\dag c_{kC}^\dag \text{ if } n_k=2 \text{ and } ( i_k, \lambda_k^3 , \lambda_k^8) = (\frac{1}{2},-\frac{1}{2},-\frac{1}{2\sqrt{3}}), \\
    &c_{kA}^\dag c_{kB}^\dag c_{kC}^\dag \text{ if } n_k=3.
\end{align*}
We also set $(P_0,Q_0,I_0,\Lambda_0^3,\Lambda_0^8)=(0,0,0,0,0)$.
Thus, a basis state such as~\eqref{eq:SDwriting} is fully determined by~$I_L$, $\Lambda_L^3$, $\Lambda_L^8$ and $P_k$, $Q_k$, $n_k$ for all $k$ from $1$ to $L$.
In Appendix~\ref{ap:exSQ}, we illustrate how to obtain the second quantization form of a state from its ssYT writing by applying Eq.~\eqref{eq:SDwriting} on some of the states in Fig.~\ref{fig:example_state}(b) and (c).  

Besides being potentially helpful from a numerical point of view, the second quantized version of the ssYT states will be useful in the next sections to characterize the effects of losses by identifying a set of conserved quantities.

\subsection{Conserved quantum numbers during loss processes} \label{sec:conservedQuantLoss}

By applying the jump operator $L_j$, defined in Eq.~\eqref{eq:Lj}, on a state characterized by the quantum numbers~$I_L$, $\Lambda_L^3$, $\Lambda_L^8$ and $\{ P_k, Q_k,n_k \}, \forall k \in \{1,\ldots, L\}$, the state is annihilated if $n_j\neq 3$ (no triple occupancy at site $j$); while if $n_j=3$ all the quantum numbers~$I_L$, $\Lambda_L^3$, $\Lambda_L^8$ and $\{ P_k, Q_k,p_k,q_k \}, \forall k \in \{1,\ldots, L\}$ are unchanged, the only modified occupation number is~$n_j$ which goes from~$3$ to~$0$. In particular, $p_j$ and $q_j$ both remain zero, as shown in Eq.~\eqref{eq:nk0or3}. We note that this result is much stronger than $[L_j,\Lambda_3]=[L_j,\Lambda_8]=[L_j,I^2]=[L_j, C_2]=[L_j, C_3]=0$ shown in Sec.~\ref{sec:SU3rot}, since the latter corresponds to the conservation of $\Lambda^3_L$, $\Lambda^8_L$, $I_L$, $P_L$ and $Q_L$ only.

\subsection{Modification rules for a ssYT state undergoing a loss} \label{sec:effect_loss_YT}

\begin{figure}[t]
\includegraphics[width=0.9\linewidth]{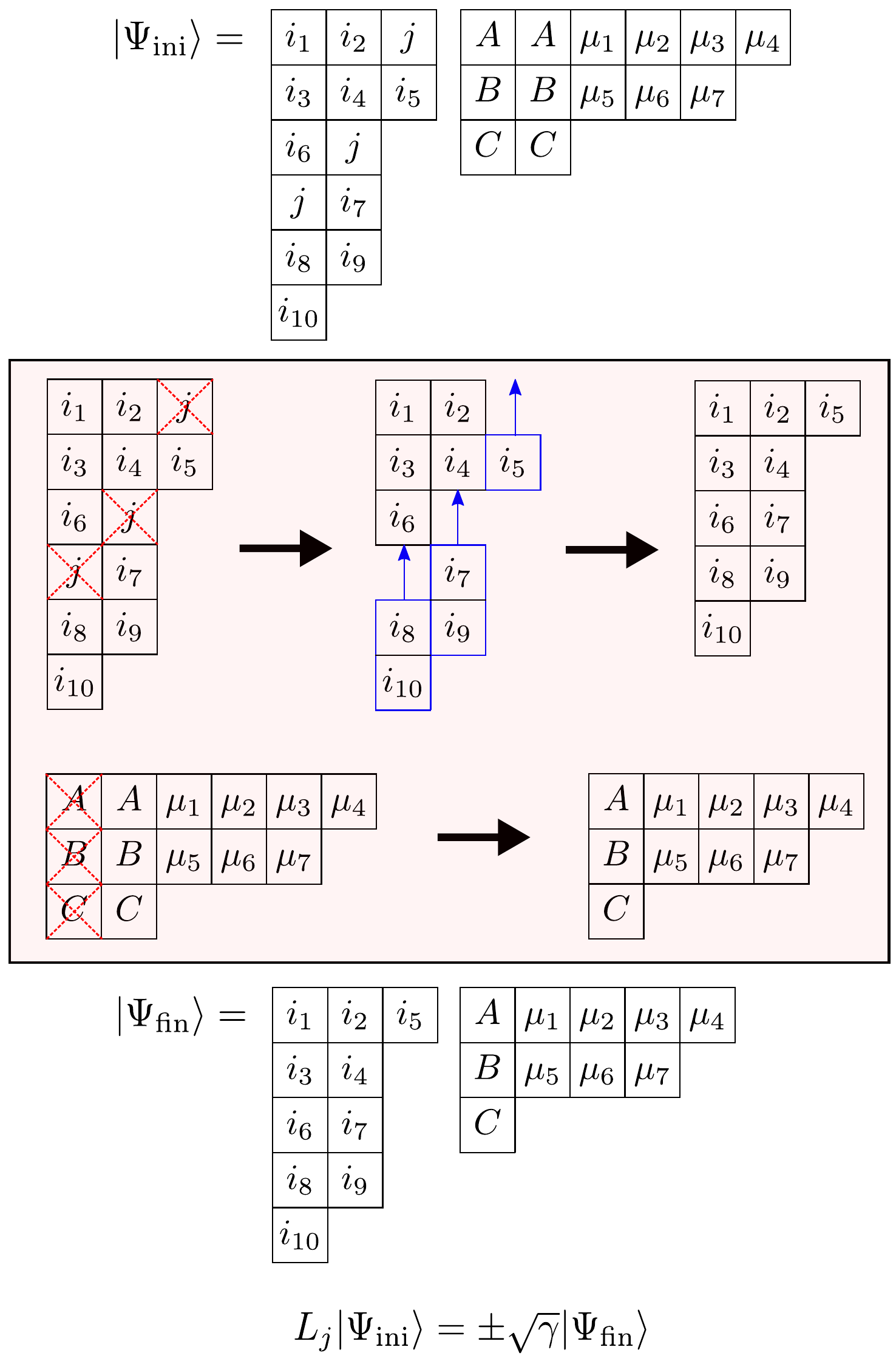}
\caption{Modification rules for a ssYT state $\ket{\Psi_{\rm ini}}$, having $n_j=3$, which undergoes a loss at site $j$.} \label{fig:effectLj}
\end{figure}

If a state does not contain exactly three particles at site~$j$ (i.e. $n_j \neq3$), it is annihilated by the jump operators~$L_j$. Otherwise, if a state contains exactly three particles at site~$j$ (i.e. $n_j =3$),  applying $L_j$ amounts to erasing the leftmost column in the spin ssYT; this column is necessarily three boxes long and contains one $A$, one $B$, and one $C$. The effect on the orbital ssYT is slightly more complex:  applying $L_j$ amounts to first removing the three boxes containing a $j$, then making the boxes below the deleted ones slide upward to form a connected tableau. This is illustrated in Fig.~\ref{fig:effectLj}.

\subsection{Direct consequence for the total number of fermions} \label{sec:Lower-Bound}
Knowing the effect of $L_j$ on the ssYT basis states provides a lower bound for the total number of fermions in the lattice, holding at any time during the dynamics.
Indeed, successively applying jump operators on a state belonging to the $(p,q)$-irrep of $\mathrm{SU}(3)$ removes the three boxes-long columns in its spin part. Thus, any state belonging to the $(p,q)$-irrep and containing exactly $\Nf = p + 2q$ fermions is non-lossy, in the sense that it is annihilated by all the jump operators. 
As a consequence, the following theorem can be stated.

\textit{Theorem} - Let $\{ \ket{\varphi_i} \}$ be an orthonormal basis of the Hilbert space diagonalizing simultaneously the operators $C_2$, $C_3$ and $\Nf$:
    \begin{equation}
    \begin{split}
        &C_2 \ket{\varphi_i} = c_2(p_i,q_i) \ket{\varphi_i}, \quad C_3 \ket{\varphi_i} = c_3(p_i,q_i) \ket{\varphi_i},\\
        &\Nf \ket{\varphi_i} = \left( p_i + 2 q_i + 3 r_i \right) \ket{\varphi_i} , \ p_i,q_i,r_i \in \{0,1,\ldots\}.        
    \end{split} \label{eq:varphi}
    \end{equation}
The mean number of particles satisfies the lower bound
    \begin{equation}
    \forall t , \ \mathcal{N}(t) := \Tr[N_{\rm f} \rho(t)] \geq  \sum_{i}  (p_i + 2 q_i) \bra{\varphi_i} \rho(0) \ket{\varphi_i} \label{eq:LowerBound_SU3}
\end{equation}
 (see Appendix~\ref{Ap:Lower-Bound} for a proof). 
 
 We notice here that this bound does of course depend on the initial state, but crucially does not depend on the precise form of the Hamiltonian governing the dynamics, as long as it is $\mathrm{SU}(3)$ invariant and on-site three-body losses are the only loss mechanism.
 
\begin{figure}[t]
\includegraphics[width=1.0\linewidth]{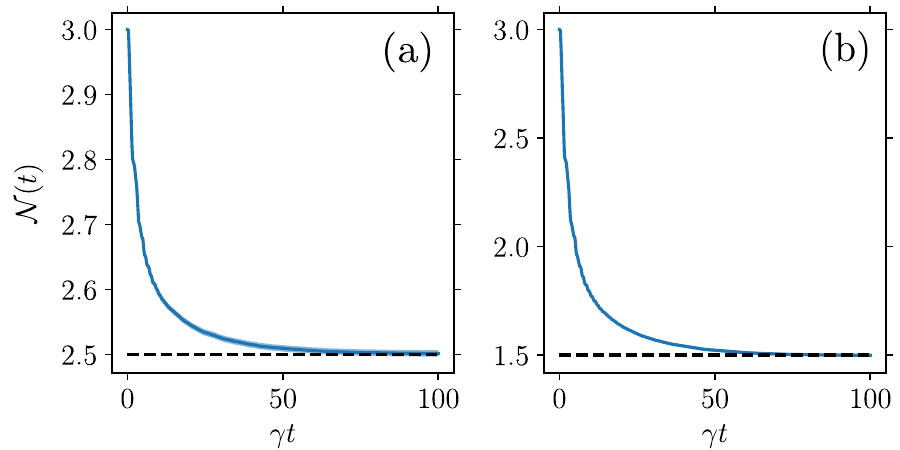}
\caption{Mean number of fermions remaining on the lattice as a function of time for the parameters $(L,J, U_2, U_3,\gamma)=(3,1,1,0,1)$ and periodic boundary conditions. (a) The initial state is $\ket{\Psi_0} = c_{1A}^\dag c_{2B}^\dag c_{3C}^\dag \ket{\rm vac}$. (b) The initial state is $\ket{\Psi_0} = \left( c_{1A}^\dag c_{2B}^\dag c_{3C}^\dag + c_{2A}^\dag c_{3B}^\dag c_{1C}^\dag + c_{3A}^\dag c_{1B}^\dag c_{2C}^\dag  \right) \ket{\rm vac}/\sqrt{3}$. For the two panels, the blue curve is computed via numerical simulations using stochastic quantum
trajectories approach with $50000$ trajectories. The black dotted line is the lower bound in Eq.~\eqref{eq:LowerBound_SU3} (see Appendix~\ref{ap:L3N3} for detailed calculation). The numerical calculation is performed with the open-source python-framework QuTiP~\cite{JOHANSSON20121760,JOHANSSON20131234}.}
\label{fig:Few_part}
\end{figure}

 In Fig.~\ref{fig:Few_part}, we show that this lower bound is saturated at late time for different initial pure states with $\Nf=3$ fermions on $L=3$ sites; in Appendix~\ref{ap:L3N3} we provide a possible basis of states $\ket{\varphi_i}$ for this particular case, and the explicit derivation of the lower bounds for the initial states considered in Fig.~\ref{fig:Few_part}.

In Eqs.~\eqref{eq:c1} and \eqref{eq:c2}, the eigenvalues of the Casimir operators, denoted $c_2$ and $c_3$, are expressed as a function as $p$ and $q$. 
It is however also possible to express $p+2q$ as a function of $c_2$ and $c_3$ as
\begin{equation}
    \begin{split}
        p+2q =  2 &\sqrt{3} \sqrt{1+c_2} \times\\ & \cos \left[  \frac{1}{3} \arccos \left( - \sqrt{3} \frac{c_3}{\left( 1+c_2 \right) ^{3/2}} \right) \right] -3, \label{eq:c1_c2_p_q}
\end{split}
\end{equation}
 which expresses the bound Eq.~\eqref{eq:LowerBound_SU3} solely in terms of the Casimir operators' eigenvalues. This highlights the importance of the $\mathrm{SU}(3)$ Casimirs in determining the final particle density of the gas, in analogy to how the squared total angular momentum plays the same role in the case of a $\mathrm{SU}(2)$ gas with local two-body losses~\cite{Foss-Feig2012,Rosso2021}.
A detailed proof of Eq.~\eqref{eq:c1_c2_p_q} can be found in Appendix~\ref{Ap:c1_c2_p_q}.

\subsection{Factorizability of the non-lossy states in first quantization} \label{sec:facto}
We now discuss the factorizability of the non-lossy states. A state is factorized between an orbital part and a spin part in first quantization language, if and only if its ssYTs writing contains only vertical and horizontal tableaux~\cite{Sakurai1993}. More precisely, if a state is of the form $\ket{\Psi} = \ket{\Psi_{\rm orb}}  \otimes \ket{\Psi_{\rm spin}}$ with 
\begin{equation}
\ket{\Psi_{\rm orb}} = 
    \begin{ytableau}
         \ & \cdots &  \
    \end{ytableau} \quad \text{and} \quad
    \ket{\Psi_{\rm spin}} = 
    \begin{ytableau}
         \ \\
          \vdots  \\
         \
    \end{ytableau}\,,
\end{equation}
then the second-quantized state is symmetric by an arbitrary exchange of two lattice sites, and antisymmetric by any exchange of two spin flavors (see Fig.~\ref{fig:example_state}(c) for examples). In the first quantization language, this means that the orbital part of the state is fully symmetric and its spin part is fully antisymmetric under the exchange of two fermions. On the other hand, if a state is such that
\begin{equation}
\ket{\Psi_{\rm orb}} =     
    \begin{ytableau}
         \ \\
        \vdots \\
         \
    \end{ytableau}\,, \quad \text{and} \quad
    \ket{\Psi_{\rm spin}} = 
    \begin{ytableau}
         \ & \cdots & \ 
    \end{ytableau}
\end{equation}
then its orbital part is fully antisymmetric and its spin part is fully symmetric.

As shown in Sec.~\ref{sec:Lower-Bound}, the non-lossy states of our $\mathrm{SU}(3)$-invariant model are in general non-factorizable in first quantization, since they can have both $p\neq 0$ and $q \neq 0$, namely they are not the product of spatial and spin wavefunctions with definite exchange symmetry.  
The feature is peculiar with respect to the case of $\mathrm{SU}(2)$-invariant Fermi gases with on-site two-body losses, which is for instance discussed in Ref.~\cite{Foss-Feig2012}. Indeed, in the latter systems the non-lossy states are expected to have a fully antisymmetric orbital part and a fully symmetric spin part. This motivates further investigation into the entanglement properties of these states.

\section{Analysis of the dark states of the lossy dynamics protected by the $\mathrm{SU}(3)$ symmetry} \label{sec:DS}

The goal of this section is to study the steady properties of the lossy dynamics and specifically to show that the $\mathrm{SU}(3)$ strong symmetry of the model allows to systematically construct families of dark states. We first show that dark states fully characterize the system when it becomes non-lossy (Sec.~\ref{sec:DSandTheorems}) and then describe how ssYTs allow to rapidly construct dark states (Sec.~\ref{Sec:DarkStates:ssYTs}).
With the help of some technical information recalled in Sec.~\ref{sec:weightD} we can enumerate the dark states appearing in different symmetry sectors (Sec.~\ref{sec:countingDS}).
Finally, we discuss some explicit examples (Secs.~\ref{sec:ferro}-\ref{subsec:pqDS}).

\subsection{The link between dark states and the late-time non-lossy dynamics} \label{sec:DSandTheorems}

We first aim to find the possible forms of the density matrix $\rho$ at late time, when the system does not undergo losses anymore. Importantly, this reduces to a simpler problem: the search of the dark states of the lossy dynamics.  
In general, a \textit{dark state} $\ket{\Psi}$ is defined as a (pure) eigenstate of the Hamiltonian which is annihilated by all the jump operators: $H_{\rm Hub} \ket{\Psi} = E \ket{\Psi}$ and $L_j \ket{\Psi} = 0, \forall j \in \{ 1, \ldots, L \}$~\cite{Kraus2008,Fazio2024}. Equivalently, a dark state is a right eigenvector of the effective non-Hermitian Hamiltonian
\begin{equation}
    H_{\rm eff} = H_{\rm Hub} - \frac{i}{2} \sum_j L_j^\dag L_j
    \label{Eq:30}
\end{equation}
with real eigenvalue (see Appendix~\ref{Ap:DSProof} for a proof).

The crucial point for linking the late-time dynamics to the dark states is that,
when the system becomes non-lossy, the density matrix takes the form
\begin{equation}
    \rho_\infty = \sum_{i,j} c_{i,j} e^{ - i \left( E_j - E_i \right) t} \ket{\Psi_j} \hspace{-0.11cm} \bra{\Psi_i}, \label{eq:rhoDS}
\end{equation}
with the coefficients $c_{i,j}$  determined by the initial conditions.
Here, each sum runs over a basis  $\{ \ket{\Psi_i} \}$ of the subspace spanned by the dark states (they form a linear space) that are eigenvectors of~\eqref{Eq:30} with real energy $E_i$. No other terms need to be included.
The proof of Eq.~\eqref{eq:rhoDS} is in Appendix~\ref{App:Eq:rhoDS}.

This explains why the study of the late-time dynamics boils down to the search for the dark states of the problem. In the rest of the paper, we will only be interested in dark states, which we will identify as right eigenvectors of Hamiltonian~\eqref{Eq:30} with real eigenvalues.

\subsection{Dark states from ssYTs}
\label{Sec:DarkStates:ssYTs}

The discussion in Sec.~\ref{sec:Effect_of_diss} provides a straightforward way of constructing dark states based on symmetries. 
The key point is that $H_{\rm eff}$ in Eq.~\eqref{Eq:30} is $\mathrm{SU}(3)$ invariant; indeed, we can interpret the non-Hermitian term as a rescaling of the three-body interaction $U_3 \to U_3 - i \gamma /2$.

The operators $\Nf$, $C_2$ and $C_3$ mutually commute and thus can be simultaneously diagonalized. 
The resulting basis is composed of states that are represented by orbital and spin ssYTs labeled by the three integers $p$, $q$, and $r$, which satisfy the following relation for the total number of particles: $\Nf = p+2q + 3 r$. Note that $p$ and $q$ are linked to $C_2$ and $C_3$ through Eqs.~\eqref{eq:c1} and~\eqref{eq:c2}, and $r \in \{0,1,2, \ldots \}$. 
Since $H_{\rm eff}$ commutes with $\Nf$, $C_2$ and $C_3$ thanks to the aforementioned $\mathrm{SU}(3)$ invariance, its eigenstates are linear combinations of ssYTs that have the same shape. In other words, the eigenvectors of $H_{\rm eff}$ have $p$, $q$, and $r$ as good quantum numbers.

Let us consider the eigenstates characterized by $r=0$: they must have real eigenvalues since by construction none of them can accommodate more than two particles on the same site.
To see that, consider the associated orbital ssYT, which is composed of 
two columns at most. By definition, the same site index cannot appear more than once in a column, and hence in this state it can at most appear twice. Therefore, these states never accommodate three particles on the same site.

As a consequence, the subspaces characterized by $\{p, q, r=0 \}$ are only composed by dark states.
Hence, by simply computing the dimensionality of these spaces, we can know the number of linearly independent states that do not undergo three-body losses because of spin-symmetry reasons.
This is what we are going to discuss in the rest of this section.
Before concluding, we mention that there also exist dark states having $r>0$ as we will see in Sec.~\ref{sec:NANBNC1}, but they are much less numerous than those with $r=0$.  

\subsection{Weight diagrams of $\mathrm{SU}(3)$} \label{sec:weightD}

We now recall the notion of weight diagram~\cite{Georgi2000}, which will be useful in the following since it will allow to count the number of dark states.

\begin{figure*}[t]
\includegraphics[width=0.9\linewidth]{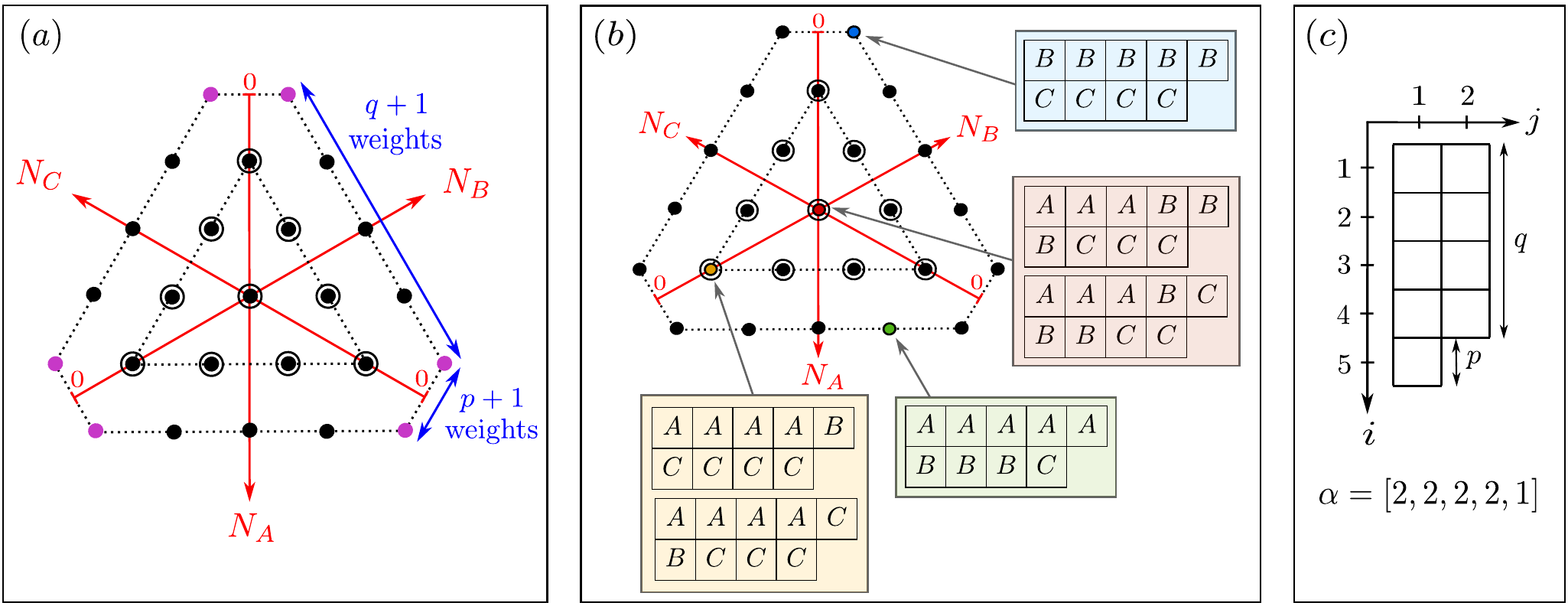}
\caption{(a)~Weight diagram for $(p=1,q=4)$. A weight space with degeneracy one is represented by a dot, while a weight space of degeneracy two is represented by a circled dot. The purple dots are weight spaces that maximize the imbalance between the three spin flavors. (b)~Relation between the weight diagram  and the spin ssYTs when $\Nf=p+2q$ for   $(p=1,q=4)$. (c)~Young diagram for the orbital part in the case $(p=1,q=4,r=0)$.} \label{fig:weightDiagram}
\end{figure*}

A \textit{weight diagram} is a way of depicting a given irrep by representing its weight spaces. For $\mathrm{SU}(3)$, a \textit{weight space} is the ensemble of spin ssYTs with fixed eigenvalues of $C_2$, $C_3$, $N_A$, $N_B$ and $N_C$; this means that both the shape and the numbers of $A$, $B$ and $C$ in the tableau is fixed in each weight space. In this framework, each spin ssYT is called a \textit{weight}. In addition, for a given $(p,q)$-irrep, the shape of the spin ssYTs is always fixed to $\overline{\alpha} = [p+q,q]$ in all the different weight spaces.
 The degeneracy of a weight space is the number of spin ssYTs it contains. In general, the weight diagram of the $(p,q)$-irrep of $\mathrm{SU}(3)$ takes the form of a hexagon where $p$ and $q$ characterize its edge lengths, as illustrated in Fig.~\ref{fig:weightDiagram}~(a). 
In this diagram, a weight space of degeneracy one is represented by a dot, while a weight space of degeneracy two is represented with a circled dot.
To obtain the weight space degeneracies, we rely on the general rules given in Ref.~\cite{Georgi2000}: the diagram is organized in layers (denoted with dashed lines in Fig.~\ref{fig:weightDiagram}~(a)),  the degeneracy of the weight spaces in the outermost layer is always one, and when we go in from a layer $\ell_1$ to the next layer $\ell_2$ the degeneracy increases by one except if $\ell_1$ has triangular shape. In Fig.~\ref{fig:weightDiagram}~(b), we illustrate the link between the weight diagram of an irrep and sets of spin ssYTs; we show the ensembles of spin ssYTs corresponding to four different weight spaces in the case $(p=1,q=4)$. 

\subsection{Counting the dark states with $\Nf=p+2q$ and $r=0$ using the Young tableaux} \label{sec:countingDS}

We now describe a method to count the number of dark states in a given sector of $N_A$, $N_B$, $N_C$, $C_2$ and $C_3$ such that $r=0$ and the total number of particles is $\Nf=p+2q$. This number of dark states, denoted $\mathcal{N}_{\rm DS}$, is equal to $\mathcal{N}_{\rm orb} \times \mathcal{N}_{\rm spin}$. Here, $\mathcal{N}_{\rm orb}$ is the total number of ssYTs of shape~$\alpha = [2^q 1^p]$ (i.e. with $q$ rows of two boxes and $p$ rows of one box) filled with integers between~$1$ and~$L$; while ~$\mathcal{N}_{\rm spin}$ is  the total number of ssYTs of shape~$\overline{\alpha} = [p+q,q]$ (i.e. with $q$ columns of two boxes and $p$ columns of one box) filled with fixed numbers of $A$, $B$ and $C$ equal to $N_A$, $N_B$ and $N_C$ respectively. The problem is thus to determine $\mathcal{N}_{\rm orb}$ and $\mathcal{N}_{\rm spin}$; to do so we will employ standard known facts from group theory.

As shown in Sec.~\ref{sec:weightD}, $\mathcal{N}_{\rm spin}$ is easily found from the weight diagram of the $(p,q)$-irrep of $\mathrm{SU}(3)$, since it is the  weight space degeneracy corresponding to $N_A$, $N_B$ and $N_C$.
For instance, for $(p=1,q=4)$ and $N_A = 4$, $N_B=3$, $N_C = 2$, we find $\mathcal{N}_{\rm spin} = 2$;  the weight diagram of the $(p=1,q=4)$-irrep is shown in Fig.~\ref{fig:weightDiagram}~(a). Remarkably, we deduce from the general structure of the weight diagrams that
\begin{equation}
\begin{split}
        \mathcal{N}_{\rm spin} = 1+  \min  \Big( p,&q, \theta,  p+q - \Theta \Big),\\
\theta = \min \left( N_A,N_B,N_C\right), &\quad \Theta = \max \left( N_A,N_B,N_C\right).
\end{split}
\end{equation}
 For completeness, we mention that $\mathcal{N}_{\rm spin}$ is the \textit{Kostka number} $K_{\lambda \mu}$ with shape~$\lambda = (p+q,q)$ and type~$\mu = (N_A,N_B,N_C)$~\cite{Stanley_Fomin_1999}.

On the other hand, $\mathcal{N}_{\rm orb}$ is given by the \textit{hook length formula} for ssYTs (also known as \textit{hook content formula})~\cite{Stanley1971}
\begin{equation}
    \mathcal{N}_{\rm orb} = \prod_{(i,j) \in \rm Y(\alpha)} \frac{L+j-i}{h_\alpha(i,j)}, \label{eq:HLFormula}
\end{equation}
where the product runs over all the boxes of the Young diagram with shape $\alpha$ denoted $\rm Y(\alpha)$; $(i,j)$ labels the box at the 
$i$th row and the $j$th column. Here, $h_\alpha(i,j)$ is the hook length of $(i,j)$ which is defined as the number of boxes $(i^\prime, j^\prime)$ such that $i^\prime = i$ and $j^\prime \geq j$ or $j^\prime = j$ and $i^\prime \geq i$. For instance, by applying Eq.~\eqref{eq:HLFormula} with $(p=1,q=4)$, we obtain $\mathcal{N}_{\rm orb} = \frac{L}{6} \frac{L+1}{4} \frac{L-1}{5} \frac{L}{3} \frac{L-2}{4} \frac{L-1}{2} \frac{L-3}{3} \frac{L-2}{1} \frac{L-4}{1}$; the corresponding Young diagram  is given in Fig.~\ref{fig:weightDiagram}~(c). 
Thus, we obtain $\mathcal{N}_{\rm DS} = (L+1)L^2(L-1)^2(L-2)^2(L-3)(L-4)/4320$ for $(p=1,q=4)$ and $N_A = 4, N_B=3, N_C = 2$. 

We stress that we have used two different techniques to compute $\mathcal{N}_{\rm spin}$ and $\mathcal{N}_{\rm orb}$ because in the former case the numbers of $A$, $B$, and $C$ fermions are constrained to be $N_A$, $N_B$, and $N_C$, respectively. On the other hand, for $\mathcal{N}_{\rm orb}$ the number of fermions at each lattice site is not fixed, so that a given integer $\{1, 2 \ldots L \}$ appears in the orbital ssYT a number of times that is not fixed.
 Note that the total number of dark states in a given $(p,q)$-sector, without fixing $N_A$, $N_B$, $N_C$, can be directly obtained as $\mathcal{N}_{\rm orb} \times d(p,q)$ where $d(p,q) := \frac{1}{2} \left( p+1 \right)  \left( q+1 \right) \left( p+q+2 \right) $ is the dimension of the $(p,q)$-irrep.

Concluding, we mention that we double-checked this analytical counting up to $L=6$ for various values of $N_A$, $N_B$ and $N_C$ by numerical exact diagonalization of the effective non-Hermitian Hamiltonian~$H_{\rm eff}$ and counting the dark states in each sector with fixed~$(p,q,r)$ (see Appendix~\ref{App:ED} for details).

\subsection{An expression for the dark states with $\Nf=p$ and $q=r=0$} \label{sec:ferro}

Remarkably, it is possible to give an expression of all the dark states in a given sector of $N_A$, $N_B$, $N_C$ and $p$ such that $\Nf=p$ (meaning $q=r=0$). The associated shape of the spin ssYTs is a single row of $p$ boxes, while the shape of the orbital ssYTs is a single column of $p$ boxes. For the rest of Sec.~\ref{sec:DS}, we assume periodic boundary conditions for simplicity, but the discussion can straightforwardly be generalized to other boundary conditions by employing the Hamiltonian's eigenmodes.

To start, we construct some simple dark states, belonging to the $(p=\Nf,q=0)$-irrep of $\mathrm{SU}(3)$.
These states will be useful for the construction of all the other ones through the action of ladder operators Eq.~\eqref{eq:Fab}.
They take the form
\begin{equation}
    \ket{\mathrm {FM}; k_1, \ldots, k_{\Nf},\sigma} = \Psi_{k_1 \sigma}^\dag \Psi_{k_2 \sigma}^\dag \ldots \Psi_{k_{\Nf} \sigma}^\dag \ket{\rm vac}, \label{eq:ferro}
\end{equation} 
with  $k_1,k_2,\ldots,k_{\Nf} \in  \frac{2 \pi}{L} \{ 1, \ldots, L \}$ and $\sigma \in \{ A,B,C\}$. 
Here, we define the Fourier modes as $\Psi_{k\sigma}^\dag = \frac{1}{\sqrt{L}} \sum_j e^{i k j} c_{j \sigma}^\dag$. The states of the form~\eqref{eq:ferro} are fully polarized; therefore, we call them $\mathrm{SU}(3)$ \textit{ferromagnetic states}. 
Thus, by the Pauli exclusion principle, they contain only single lattice site occupancies. Therefore, they are annihilated by both the local three-body loss jump operators $L_j$, and the local two-body and three-body interaction Hamiltonians $H_{\rm int,2}$ and $H_{\rm int,3}$. Moreover, these states are eigenstates of the hopping Hamiltonian which can be rewritten as
    \begin{equation}
    H_{\rm hop} = \sum_{k,\sigma} \epsilon_k \Psi_{k\sigma}^\dag  \Psi_{k\sigma} \quad \text{with} \quad \epsilon_k = - 2 J \cos k.
\end{equation}
Thus, they are trivially dark and their energies are $E = \epsilon_{k_1} + \ldots + \epsilon_{k_{\Nf}}$. Since a one-particle state $c_{j \sigma}^\dag \ket{\rm vac}$ belongs to the fundamental irrep $(p=1,q=0)$, the state~\eqref{eq:ferro} consisting of ${\Nf}$ delocalized fermions belongs to the $(p={\Nf},q=0)$-irrep of $\mathrm{SU}(3)$. 

However, other dark states with arbitrary $N_A$, $N_B$, and $N_C$ can be obtained by applying ladder operators $F_{\alpha \beta}$, defined in Eq.~\eqref{eq:Fab}, with $\alpha \neq \beta$ on the states~\eqref{eq:ferro}. The new states constructed in this way are indeed dark because the ladder operators commute with both the full Hamiltonian $H_{\rm Hub}$ and the operator $\sum_j n_{jA} n_{jB} n_{jC}$ counting the number of triple occupancies.  Since  the ladder operators also commute with both $C_2$, $C_3$ and $\Nf$, the new dark states still remain in the $(p=\Nf,q=0)$-irrep of $\mathrm{SU}(3)$, and the number of particles remains $\Nf$.

\begin{figure}[t]
\includegraphics[width=\linewidth]{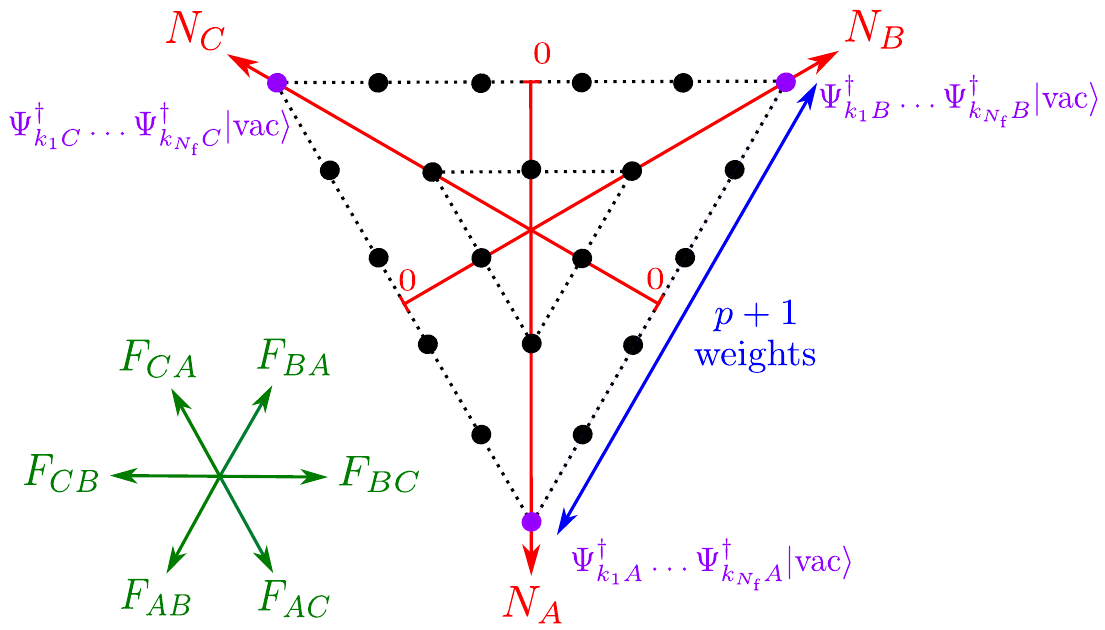}
\caption{Weight diagram of the $(p=5,q=0)$-irrep of $\mathrm{SU}(3)$. The dark states belonging to this irrep such that $\Nf = p =5$ can be arranged as the weight diagram. In this representation, each dot corresponds to a given state, and the states of Eq.~\eqref{eq:ferro} are on the three vertices of the triangle.} \label{fig:weightDiagram2}
\end{figure}

By fixing $\Nf$ distinct momenta $k_1$, $k_2$, $\ldots$, $k_{\Nf}$ among the $L$ that are possible and successively applying the ladder operators on~\eqref{eq:ferro}, we thus construct a family of dark states.  The latter can be arranged as the weight diagram of the $(p=\Nf,q=0)$-irrep of $\mathrm{SU}(3)$, i.e., in the shape of a triangle pointing down with edge length $p+1$. In this representation, the three states~\eqref{eq:ferro} with $\sigma = A,B,C$ are at the three vertices of this triangle, as shown in Fig.~\ref{fig:weightDiagram2}. Applying a ladder operator is equivalent to moving from a weight to a neighboring one in the triangle. Since, for a triangle, the weight diagram does not contain any degenerate weight space, there exists only one state for fixed $N_A$, $N_B$, $N_C$, $k_1$, $\ldots$, $k_{\Nf}$. Therefore, when only $N_A$, $N_B$, and $N_C$ are fixed and the momenta can freely vary, there exist
\begin{equation}
\mathcal{N}_{\rm DS} = \binom{L}{\Nf}    
\end{equation}
dark states which can be constructed with this method. On the other hand, by using the counting method described in Sec.~\ref{sec:countingDS}, we obtain exactly the same number of dark states, which guarantees us that we have found all the dark states with $p\neq0$ and $q=r=0$.  

\subsection{Three-color $\eta$-pairing dark states having $\Nf=2q$ and $p=r=0$}
Some dark states consisting in delocalized doublons are known as \textit{three-color $\eta$-pairing states}~\cite{Nakagawa2024}, which are a generalization of the $\eta$-pairing states in the $\mathrm{SU}(2)$ Hubbard model~\cite{Yang1989}. 
In the periodic boundary condition case, the number of sites $L$ should be even; no such constraint is necessary with open boundaries.
In general, a three-color $\eta$-pairing state, containing $N_{\rm AB}$, $N_{\rm AC}$, and $N_{\rm BC}$ delocalized doublons $A$-$B$, $A$-$C$, and $B$-$C$, respectively, is written as
\begin{equation}
   \ket{\eta; N_{\rm AB}, N_{\rm AC},N_{\rm BC}} = (\eta_{\rm AB}^\dag)^{N_{\rm AB}}  (\eta_{\rm AC}^\dag)^{N_{\rm AC}}  (\eta_{\rm BC}^\dag)^{N_{\rm BC}} \ket{\rm vac} \label{eq:eta}
\end{equation}with $N_{\rm AB}, N_{\rm AC}, N_{\rm BC} \in \{ 0,1,2,\ldots \}$. Here, we have introduced the $\eta$-pairing operators:
\begin{equation}
 \eta_{\rm \alpha \beta}^\dag = \sum_{j} (-1)^j c_{j \alpha}^\dag c_{j \beta}^\dag = \sum_k \Psi_{k \alpha}^\dag \Psi_{\pi-k \beta}^\dag \label{eq:etaOP}
\end{equation}
with $\alpha,\beta \in \{ A,B,C\}$ and $\alpha \neq \beta$.
The states of the form~\eqref{eq:eta} have only double occupancies, since two doublons (e.g. $AB$ and $AC$) cannot be accommodated on the same site by Pauli exclusion principle. This means that the class of states in Eq.~\eqref{eq:eta} are annihilated by all the jump operators.
They are actually eigenvectors of the Hamiltonian $H_{\rm Hub}$ and thus are dark states, because they satisfy
\begin{subequations} \label{eq:etaEV}
\begin{equation}
         H_{\rm hop} \ket{\eta; N_{\rm AB}, N_{\rm AC},N_{\rm BC}} = 0,
\end{equation}
\begin{equation}
\begin{split}
        &H_{\rm int,2}   \ket{\eta; N_{\rm AB}, N_{\rm AC},N_{\rm BC}} \\&= U_2 \left(  N_{\rm AB}  + N_{\rm AC} + N_{\rm BC} \right)  \ket{\eta; N_{\rm AB}, N_{\rm AC},N_{\rm BC}},
\end{split}
\end{equation}
\begin{equation}
         H_{\rm int,3} \ket{\eta; N_{\rm AB}, N_{\rm AC},N_{\rm BC}} = 0,
\end{equation} 
\end{subequations}
(see Appendix~\ref{Ap:eta} for a proof).
  
Since a doublon of the form $c_{j\alpha}^\dag c_{j \beta}^\dag \ket{\rm vac}$ belongs to the anti-fundamental irrep $(p=0,q=1)$, the state~\eqref{eq:eta} belongs to the $(p=0, q=N_{\rm AB} + N_{\rm BC} + N_{\rm AC})$-irrep of $\mathrm{SU}(3)$ and thus satisfies the relation for the total number of fermions $\Nf= 2 N_{\rm AB} + 2 N_{\rm BC} + 2 N_{\rm AC}=2q$.
Additionally, applying a ladder operator $F_{\alpha \beta}$ on the state~\eqref{eq:eta} creates a state of the same form; only the numbers $N_{\rm AB}$, $N_{\rm AC}$ and $N_{\rm BC}$ are modified. For instance, we have
 \begin{equation}
     \begin{split}
              F_{\rm AB} &(\eta_{\rm AB}^\dag)^{N_{\rm AB}} (\eta_{\rm AC}^\dag)^{N_{\rm AC}} (\eta_{\rm BC}^\dag)^{N_{\rm BC}}  \ket{\rm vac}\\ 
           &= (\eta_{\rm AB}^\dag)^{N_{\rm AB}} (\eta_{\rm AC}^\dag)^{N_{\rm AC}+1}  (\eta_{\rm BC}^\dag)^{N_{\rm BC}-1} \ket{\rm vac}.          
     \end{split}
 \end{equation}
This can be shown from the commutation relations
\begin{equation}
    [\eta_{\alpha \beta}^\dag , F_{\gamma \delta}] = \delta_{\alpha,\delta} \eta_{\beta \gamma}^\dag - \delta_{\beta,\delta} \eta_{\alpha \gamma}^\dag.
\end{equation}

\begin{figure}[t]
\includegraphics[width=\linewidth]{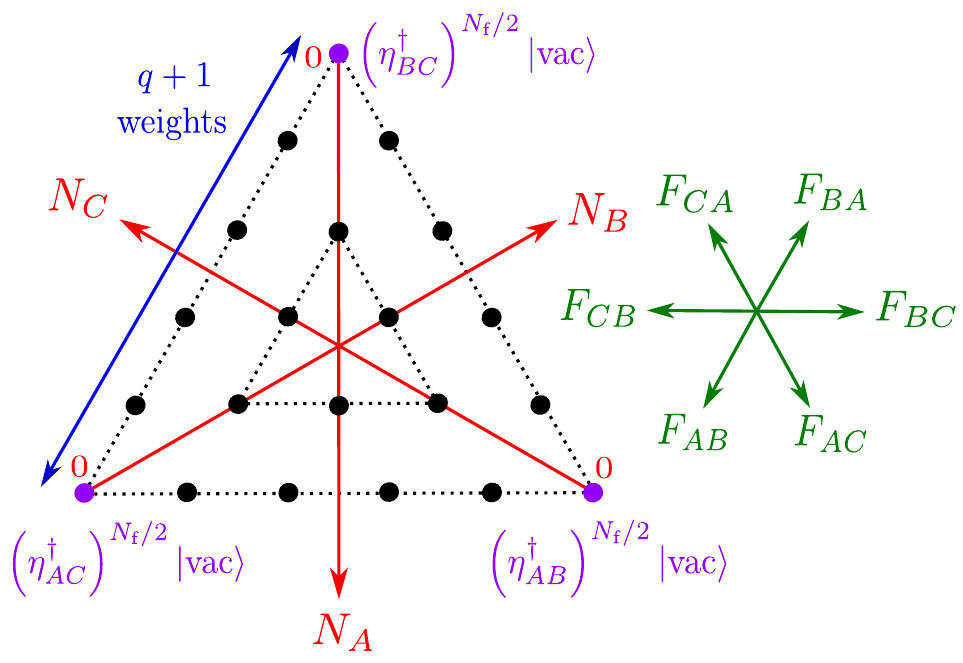}
\caption{Weight diagram for the $(p=0,q=5)$-irrep of $\mathrm{SU}(3)$. The three-color $\eta$-pairing dark states belonging to this irrep, i.e satisfying $\Nf=2q =10$, can be arranged as the weight diagram.} \label{fig:weightDiagram2bis}
\end{figure}

In a similar way as in Sec.~\ref{sec:ferro}, the states of the form~\eqref{eq:eta}, such that the particle number $\Nf$ is fixed, can be arranged as the weight diagram of the $(p=0,q=\Nf/2)$-irrep of $\mathrm{SU}(3)$, i.e., in the shape of a triangle pointing up with edge length $q+1$. In this representation the vertices of the triangle correspond to the states $\left( \eta_{AB}^\dag \right)^{\Nf/2} \ket{\rm vac}$, $\left( \eta_{AC}^\dag \right)^{\Nf/2} \ket{\rm vac}$ and $\left( \eta_{BC}^\dag \right)^{\Nf/2} \ket{\rm vac}$, as shown in Fig.~\ref{fig:weightDiagram2bis}. 

However, the $\eta$-pairing states~\eqref{eq:eta} do not exhaust all the possible dark states in the $(p=0,q=\Nf/2)$-irrep. Indeed, we predict the existence of only one $\eta$-pairing dark state for fixed numbers of $A$, $B$ and $C$ fermions 
because once these values are fixed, so are the values of $N_{\rm AB}$, $N_{\rm AC}$, $N_{\rm BC}$: the counting method presented in Sec.~\ref{sec:countingDS} predicts the existence of many more dark states in this irrep.
Although we did not investigate this problem further,
we expect that a general dark state with $p=0$ and $q\neq0$ is also composed of singlets of the form $\left( c_{i \alpha}^\dag c_{j \beta}^\dag + c_{j \alpha}^\dag c_{i \beta}^\dag \right) \ket{\rm vac}$.

\subsection{Dark states having $\Nf=p+2q$ and $r=0$ with  $p,q\neq 0$} \label{subsec:pqDS}

Additional dark states are constructed by successively applying Fourier mode operators $\Psi_{k\sigma}^\dag = \frac{1}{\sqrt{L}} \sum_j e^{i k j} c_{j \sigma}^\dag$ and $\eta$-pairing operators~\eqref{eq:etaOP} on the vacuum state:
    \begin{equation}
    \ket{{\rm\eta FM}; \alpha,\beta, N_{\alpha \beta},k_1,\ldots,k_n} = (\eta_{\rm \alpha \beta}^\dag)^{N_{\rm \alpha \beta}} \Psi_{k_1 \alpha}^\dag \ldots \Psi_{k_n \alpha}^\dag \ket{\rm vac} \label{eq:mixed}
\end{equation}
with $\alpha,\beta \in \{ A,B,C\}$, $\alpha \neq \beta$, and $k_1, \ldots,k_n \in \frac{2 \pi}{L} \{1, \ldots, L \}$. 
Here, $n$ is the number of times an operator of the form $\Psi^\dagger_{k \sigma}$ has been applied and $2 N_{\alpha \beta}$ the number of $\eta$-paired fermions, meaning that the state~\eqref{eq:mixed} contains $\Nf = n + 2 N_{\alpha \beta}$ fermions in total. This state satisfies
\begin{subequations}
\begin{equation}
\begin{split}
        &H_{\rm hop} \ket{{\rm\eta FM}; \alpha,\beta, N_{\alpha \beta},k_1,\ldots,k_n} \\&= \left( \epsilon_{k_1} + \ldots + \epsilon_{k_n} \right) \ket{{\rm\eta FM}; \alpha,\beta, N_{\alpha \beta},k_1,\ldots,k_n},
\end{split}
\end{equation}
\begin{equation}
\begin{split}
        &H_{\rm int,2} \ket{{\rm\eta FM}; \alpha,\beta, N_{\alpha \beta},k_1,\ldots,k_n}\\ &= U_2 N_{\rm \alpha \beta}\ket{{\rm\eta FM}; \alpha,\beta, N_{\alpha \beta},k_1,\ldots,k_n},
\end{split}
\end{equation}
\begin{equation}
    H_{\rm int,3} \ket{{\rm\eta FM}; \alpha,\beta, N_{\alpha \beta},k_1,\ldots,k_n} = 0.
\end{equation}
\end{subequations}

The state~\eqref{eq:mixed} only contains two different spin components $\alpha$ and $\beta$, thus it is annihilated by all the three-body loss jump operators. We note that a state of the form
$(\eta_{\rm A B}^\dag)^{N_{\rm A B}} \Psi_{k_1 C}^\dag \ldots \Psi_{k_n C}^\dag \ket{\rm vac}$
cannot be a dark state because it contains triple occupancies. 
Since the state~\eqref{eq:mixed} contains $n$ delocalized plane-wave states and $N_{\alpha \beta}$ delocalized $\eta$-paired doublons, it belongs to the $(p=n, q=N_{\alpha \beta})$-irrep of $\mathrm{SU}(3)$. Therefore, it satisfies the relation $\Nf=p+2q$. 

\begin{figure}[t]
\includegraphics[width=\linewidth]{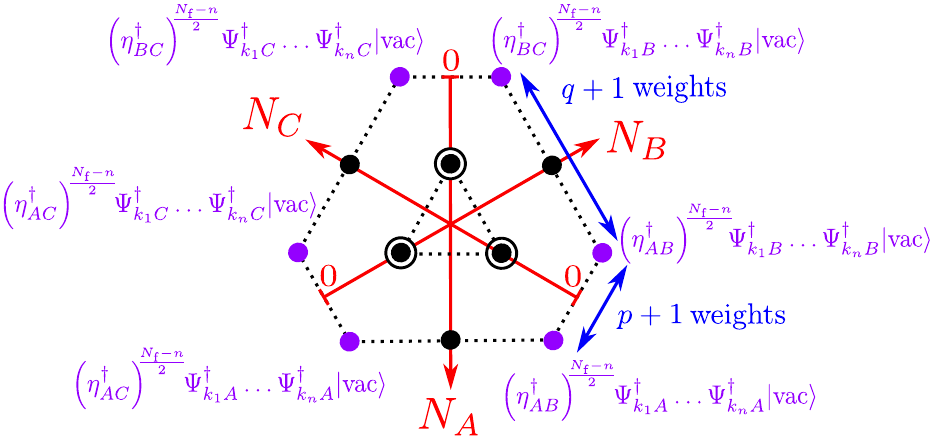}
\caption{Weight diagram for the $(p=n=1,q=(\Nf-n)/2=2)$-irrep of $\mathrm{SU}(3)$. The dark states belonging to this irrep, i.e. satisfying $\Nf = p+2q$ can be arranged as the weight diagram.} \label{fig:weightDiagrambis}
\end{figure}

Once again, by successively applying the ladder operators~\eqref{eq:Fab} on the state~\eqref{eq:mixed}, we generate a family of dark states belonging to the $(p=n, q=N_{\alpha \beta})$-irrep of $\mathrm{SU}(3)$. This set of dark states can be arranged as the weight diagram of the $(p=n, q=N_{\alpha \beta})$-irrep, i.e., in the shape of a hexagon, as illustrated in Fig.~\ref{fig:weightDiagrambis}. In this case, degeneracies appear in the weight diagram. Thus, the multiplicity of a weight space corresponds to the number of independent dark states, with fixed numbers of $A$, $B$, and $C$ fermions, which can be constructed from a given state of the form~\eqref{eq:mixed}. Also here, the states of the form~\eqref{eq:mixed} do not exhaust all the possible dark states in the $(p,q)$-irrep with both $p \neq 0$ and $q \neq 0$.

\section{Dark states for $N_A = N_B = N_C = 1$} \label{sec:NANBNC1}

In this section, we discuss the dark states in the case of $N_A = N_B  = N_C = 1$. Since the total number of fermions satisfies $\Nf = 3=p + 2 q +3r$, we know that dark states are present in the sectors $(p=3,q=0,r=0)$ and $(p=1,q=1,r=0)$.  
But, as we will see, there also exist non-trivial dark states in the sector $(p=0,q=0,r=1)$ appearing because of subtle quantum-interference effects that are not directly related to the $\mathrm{SU}(3)$ symmetry.

\subsection{Dark states protected by the $\mathrm{SU}(3)$ symmetry}

\begin{figure}[t]
\includegraphics[width=0.8\linewidth]{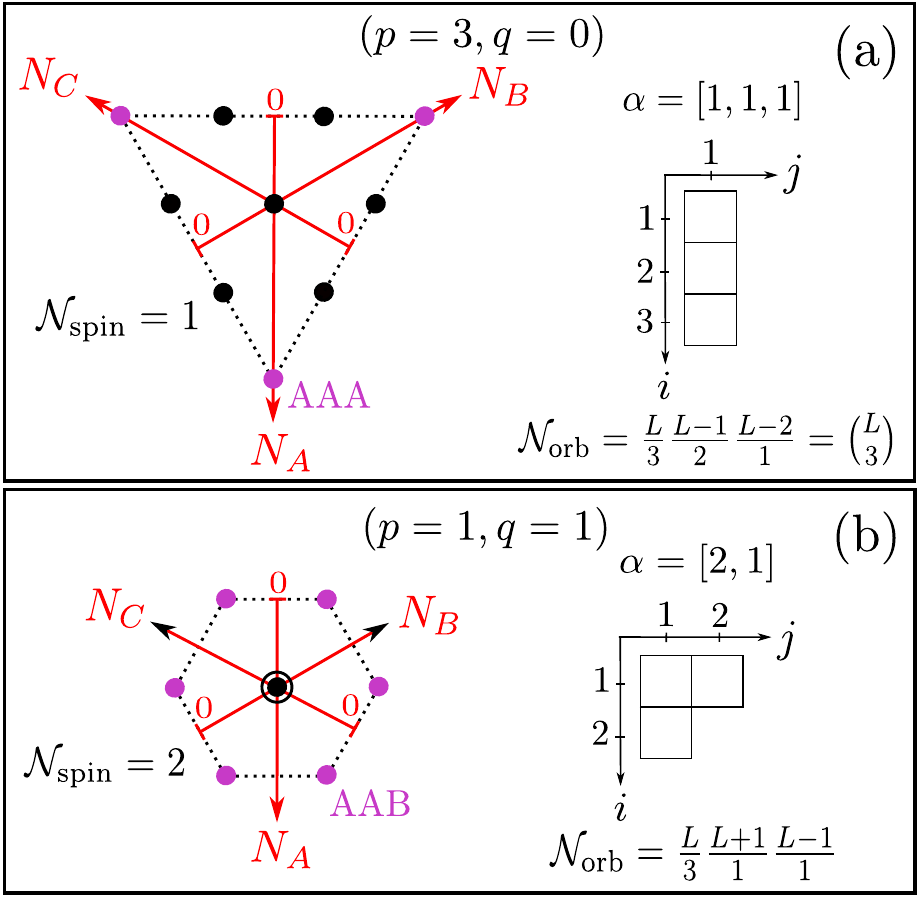}
\caption{Weight diagrams for the spin part and Young diagrams for the orbital part in the case of three particles; in the sectors $(p=3,q=0)$ (panel (a)) and $(p=1,q=1)$ (panel (b)). $\mathcal{N}_{\rm orb}$ is determined by using Eq.~\eqref{eq:HLFormula};  for $N_A=N_B=N_C=1$,  $\mathcal{N}_{\rm spin}$ is the degeneracy of the weight space at the center of the weight diagram. The number of dark states in each sector is $\mathcal{N}_{\rm orb} \times \mathcal{N}_{\rm spin}$.} \label{fig:weightDiagram3}
\end{figure}

The results on the calculations of the dark states protected by the $\mathrm{SU}(3)$ symmetry are illustrated in Fig.~\ref{fig:weightDiagram3}. 
By carefully using the counting of ssYTs explained in Sec.~\ref{sec:countingDS}, we find that there are $\binom{L}{3}$ dark states in the sector $(p=3,q=0,r=0)$; in fact, $\mathcal{N}_{\rm spin} = 1$ and $\mathcal N_{\rm orb} = \binom{L}{3}$.
When we consider the symmetry sector $(p=1,q=1,r=0)$, the number of dark states is
$2 \frac{L}{3} (L+1)(L-1)$, with $\mathcal{N}_{\rm spin} = 2$ and $\mathcal N_{\rm orb} = \frac L3 (L^2-1)$. 
To get a better grasp on the dark states that we just found,
we now describe a method to find a basis of these $\mathrm{SU}(3)$-protected dark subspaces, for which $\Nf=p+2q$, without using any knowledge about Young tableaux. 

We start by constructing a basis of states corresponding to a corner of the weight diagram. We define a corner as a weight space with maximal imbalance between the three spin components $A$, $B$ and $C$; the corners are drawn with purple dots in Figs.~\ref{fig:weightDiagram} to~\ref{fig:weightDiagram3}. Then, we apply the ladder operators~\eqref{eq:Fab} on this set of states to reach the center of the weight diagram ($N_A = N_B=N_C =1$). There are many different paths in the weight diagram to achieve this, but the number of paths that produce independent states starting from the same state at a given corner equals the degeneracy of the targeted weight space.

For instance, in the sector $(p=3,q=0)$, we choose the corner at the bottom of the triangle, denoted $AAA$ in Fig.~\ref{fig:weightDiagram3}~(a). A basis of the dark subspace corresponding to this corner is the set of states $c_{i A}^\dag c_{j A}^\dag c_{l A}^\dag \ket{\rm vac}$ with $i$, $j$ and $l$ three distinct lattice sites varying between $1$ and $L$. This basis contains $\binom{L}{3}$ states as expected. Moreover, a basis of the dark subspace corresponding to the center of the weight diagram is composed of the states
\begin{equation}
    F_{CA}F_{BA}c_{i A}^\dag c_{j A}^\dag c_{l A}^\dag \ket{\rm vac}.
\end{equation}

 In the sector $(p=1,q=1)$, we choose the corner at the bottom-right of the hexagon, denoted $AAB$ in Fig.~\ref{fig:weightDiagram3}~(b). A basis of the dark subspace corresponding to this corner is composed of states of the form $c_{i A}^\dag c_{iB}^\dag c_{jA}^\dag \ket{\rm vac}$ and states of the form $\left( c_{i A}^\dag c_{jB}^\dag + c_{j A}^\dag c_{i B}^\dag  \right) c_{lA}^\dag \ket{\rm vac}$ with $i$, $j$ and $l$ distinct. There are $L(L-1)$ states of the first form. For the second form of state, the pair of sites $(i,j)$ can take $L(L-1)/2$ values, while $l$ can take $(L-2)$ values $\textit{a priori}$. In fact, the number of independent states of the second form is $2/3 \times L(L-1)/2 \times(L-2)$ since among the three states
 \begin{equation}
     \begin{split}
         &\left( c_{i A}^\dag c_{jB}^\dag + c_{j A}^\dag c_{i B}^\dag  \right) c_{lA}^\dag \ket{\rm vac},\\
         &\left( c_{j A}^\dag c_{lB}^\dag + c_{l A}^\dag c_{j B}^\dag  \right) c_{iA}^\dag \ket{\rm vac},\\
         &\left( c_{i A}^\dag c_{lB}^\dag + c_{l A}^\dag c_{i B}^\dag  \right) c_{jA}^\dag \ket{\rm vac}, \\
     \end{split}
 \end{equation}
 only two are linearly independent. Therefore, a basis of the dark subspace corresponding to the center of the weight diagram for $(p=1,q=1)$ is composed of the states 
 \begin{equation}
     \begin{split}
         &F_{CA} c_{i A}^\dag c_{iB}^\dag c_{jA}^\dag \ket{\rm vac}, \\
         &F_{CB} F_{BA} c_{i A}^\dag c_{iB}^\dag c_{jA}^\dag \ket{\rm vac}, \\
         &F_{CA} \left( c_{i A}^\dag c_{jB}^\dag + c_{j A}^\dag c_{i B}^\dag  \right) c_{lA}^\dag \ket{\rm vac}, \\
         &F_{CB} F_{BA} \left( c_{i A}^\dag c_{jB}^\dag + c_{j A}^\dag c_{i B}^\dag  \right) c_{lA}^\dag \ket{\rm vac}.
     \end{split}
 \end{equation}
Finally, the dimension of this dark subspace is $2 \times L(L-1) + 2\times 2/3 \times L(L-1)/2 \times(L-2)= 2/3 \times L (L+1)(L-1)$ as expected.

\subsection{Dark states that are not protected by the $\mathrm{SU}(3)$ symmetry}

So far, we have considered subspaces for fixed~$p$ and~$q$ such that $r=0$. In these cases, the entire subspace is dark. However, it turns out that the sector $(p=0,q=0,r=1)$ contains both states that can dissipate and dark states. 

In the sector $(p=0,q=0,r=1)$, the states can be decomposed on a basis of ssYTs for which the Young diagram for the spin part is a single column of three boxes, while the Young diagram for the orbital part is a single row of three boxes. Thus, as discussed in Sec.~\ref{sec:facto}, the states are factorized in first quantization as $\ket{\Psi} = \ket{\Psi_{\rm orb}} \otimes \ket{\Psi_{\rm spin}}$; the spin part is fully antisymmetric by any exchange of two fermions. Therefore, the spin part can only be of the form
\begin{equation}
\begin{split}
        \ket{\Psi_{\rm spin}} = \ket{ABC}& - \ket{BAC} + \ket{CAB}\\ - &\ket{ACB} + \ket{BCA} - \ket{CBA} = \begin{ytableau}
            A \\
            B \\
            C
        \end{ytableau}.
\end{split}
\end{equation}

On the other hand, the orbital part of the state $\ket{\Psi_{\rm orb}}$ is fully symmetric by any exchange of two fermions;
the calculation of orbital properties can be done via the study of the properties of fake spinless bosons.
Specifically, the action of the Fermi-Hubbard Hamiltonian $H_{\rm Hub}$ reduces to that of the Bose-Hubbard Hamiltonian
\begin{equation}
    H_{\rm Hub}^B = -J \sum_{j=1}^L \left( b_{j+1}^\dag b_j + \rm{h.c} \right)+\frac{U_2}{2} \sum_{j=1}^L n_{j}(n_j-1)
\end{equation}
on the orbital part of the states, which is now rewritten in terms of bosons. Here, $b_j^\dag$ and $b_j$ denote the creation and annihilation operators of a boson at site $j$, and $n_j = b_j^\dag b_j$ is the local density operator for bosons. In order to find the fermionic dark states in the sector $(p=0,q=0,r=1)$, we study this simpler bosonic system. 
The identification of the dark states that are of our interest simply demands the search for three-body eigenstates of $H_{\rm Hub}^B$ having no triple lattice site occupancy. We mention that, in dimension $d > 1$ and with periodic boundary conditions, the many-body eigenstates of the Bose-Hubbard Hamiltonian with no double lattice site occupancy have been studied in Ref.~\cite{Suto2002}.

In the following, we use the notation $$\ket{ i \ j \ l} := b_{i}^\dag  b_{j}^\dag  b_{l}^\dag \ket{\rm vac}.$$ An exact diagonalization of $H_{\rm Hub}^B$ can be performed with the open-source Python package QuSpin~\cite{Weinberg2017,Weinberg2019}. With this numerical method, we find that there exists only one dark state for open boundary conditions:
\begin{equation}
    \ket{2 \ 2 \ 1}  - \ket{2 \ 2 \ 3} - 2 \ket{ 1 \ 1 \ 3 } + 2 \ket{ 3 \ 3 \ 1} \text{ with } L=3.
\end{equation}
This state is annihilated by the hopping part of the Hamiltonian and has a well-defined interaction energy~$U_2$. For $L\neq3$, no dark states are found with open boundary conditions. However, for periodic boundary conditions, we observe that there exist dark states for any system size $L \geq 3$, and none for $L<3$. In Table~\ref{tab:DS}, we give the dark states for $L=3,4,5$. 

For general $L$, we can identify a closed non-dissipative subspace generated by the states:
\begin{widetext}
\begin{subequations}
    \begin{align}
                \ket{\Psi_\Delta^\delta} =& \sum_{j=1}^L e^{i \delta j} \left( \ket{j \ j \ j+\Delta}  -  \ket{j-\Delta \ j \ j} \right), \qquad \forall \Delta \geq 1; \label{eq:PsiD} \\ 
                \ket{\Phi_\Delta^\delta} =& \sum_{j=1}^L e^{i \delta j} ( \ket{j \ j+1 \ j+\Delta}  -  \ket{j-\Delta \ j-1 \ j} ), \qquad \forall \Delta \geq 3; \\
                \ket{\Gamma_{\Delta_1 \Delta_2}^\delta} =& \sum_{j=1}^L e^{i \delta j} ( \ket{j \ j+\Delta_1 \ j+\Delta_2}  -  \ket{j-\Delta_2 \ j-\Delta_1 \ j} ),   \qquad  \forall \Delta_1 \geq 2 \quad \forall \Delta_2 \geq \Delta_1+2.
    \end{align} \label{eq:DS_generators}
\end{subequations}
\end{widetext}
In all three families of states, $\delta=0$ if $L$ is odd, while $\delta \in \{0,\pi \}$ if $L$ is even, because the states should be invariant by translation of $L$ sites. In Appendix~\ref{Ap:3bodyDSboson}, we show that this subspace is indeed closed under Hamiltonian evolution and its dimension is growing as $L^2$ for large system size. An exact analytical formula for the space dimension is given in Appendix~\ref{Ap:3bodyDSboson}. We have also checked, for $L \in \{ 5 ,\ldots, 25\}$, that the dimension of this non-dissipative subspace is equal to the total number of dark states in the sector $(p=0,q=0,r=1)$ which is found numerically by exact diagonalization of $H_{\rm Hub}^B$.
\begin{table}[h!]
\begingroup
\renewcommand{\arraystretch}{2}
\begin{tabular}{l||c}
   $L$ & \shortstack{Dark states and their energies} \\
   \hline \hline
    $3$ & $\ket{\Psi_1^{\delta=0}}$, $E = 3J+U_2$ \\[0.1cm]
 \hline
  $4$ &  $\ket{\Psi_1^{\delta=0}}$, $E=U_2+2J$ and $\ket{\Psi_1^{\delta=\pi}}$, $E=U_2-2J$  \\[0.1cm]
  & $\ket{112}- \ket{114}+\ket{332}-\ket{334}+2\ket{442}-2\ket{224}$, $E=U_2$ \\[0.1cm]
   & $\ket{221}- \ket{223}+\ket{441}-\ket{443}+2\ket{113}-2\ket{331}$, $E=U_2$ \\[0.1cm]
  \hline
  $5$ &  $\ket{\Psi_1^{\delta=0}}+ \Phi_{\pm} \ket{\Psi_2^{\delta=0}}$, $E_{\pm}= J(2-\Phi_{\pm})+U_2$, \\ & with $\Phi_{\pm} = \frac{1 \pm \sqrt{5}}{2}$ \\[0.1cm]
\end{tabular}
\endgroup
\caption{Three-body dark states of the bosonic problem i.e. eigenstates of $H_{\rm Hub}^B$ without triple occupancy, for small system size $L\in \{3,4,5\}$. These dark states are found by exact diagonalization of $H_{\rm Hub}^B$. We use the notation $\ket{ijk} := b_{i}^\dag  b_{j}^\dag  b_{k}^\dag \ket{\rm vac}$ and $\ket{\Psi_\Delta^\delta}$ is defined in Eq.~\eqref{eq:PsiD}. For $L=4$, we note that two dark states do not belong to the subspace generated by the states of Eqs.~\eqref{eq:DS_generators}.}
\label{tab:DS}
\end{table}

\section{Discussion}
\label{Sec:Discussion}

We conclude the article by discussing the generality of the results we just presented.
Most of the results of this paper are tightly linked to the existence of an $\mathrm{SU}(3)$ strong symmetry of the Lindblad dynamics, and thus do not depend on whether we study a one-dimensional or a higher-dimensional lattice. 
Starting from Sec.~\ref{sec:symmetries}, where we discussed the $\mathrm{SU}(3)$ group as a strong symmetry of the model, we observe that dimensionality never plays a role. 
Specifically, when discussing ssYTs in Sec~\ref{sec:YT}, for the orbital part we just need to label all lattice sites with an integer, which can be done for any Bravais lattice, no matter its dimensionality.
Sec.~\ref{sec:Effect_of_diss}, where we describe losses from the viewpoint of ssYTs, is also very general and its results do not depend on the geometry of the lattice.
Even the beginning of Sec.~\ref{sec:DS}, where we count the dark states and link them to the non-lossy dynamics, can be easily re-employed in a higher-dimensional lattice.

It is also to be stressed that translational invariance is not related to the $\mathrm{SU}(3)$ symmetry and thus one could also consider disordered Hamiltonians and/or disordered jump operators: the results that are valid in higher-dimensional lattices would hold true also in this case.
This of course means that they are valid also in the presence of a harmonic confinement, which is rather widespread in cold-atom setups.

The $\mathrm{SU}(3)$ symmetry is also present when considering a quantum gas in the continuum that is not trapped on a lattice.
In fact, the results in Sec.~\ref{sec:symmetries} extend also to this case. The treatment of the ssYTs for the orbital part of the wavefunction requires a bit of care because it is necessary to employ a discrete basis, like that of plane waves with quantization imposed by boundary conditions. Differently from what is discussed here, the basis is infinite-dimensional, but we expect it not to be a problem.
On the other hand, the results in Sec.~\ref{sec:Effect_of_diss} need to be completely reformulated since losses are non-local in this basis of delocalized states; this is left for future work.

Finally, we stress that the results presented starting from Sec.~\ref{sec:ferro} should instead be adapted to different lattices and we highlight that Sec.~\ref{sec:NANBNC1} is specific to one-dimensional lattices without disorder.
Assessing how the three-particle dark subspaces, described in Sec.~\ref{sec:NANBNC1}, are modified or destroyed by increased dimensionality or by external confinement is something left for future work.

%
%
\section{Conclusions} \label{sec:conclu}

In this article, we have shown that an $\mathrm{SU}(3)$-symmetric Fermi-Hubbard model with local three-body losses exhibits eight independent strong symmetries, 
which correspond to the generators of the $\mathrm{SU}(3)$ pseudo-spin algebra. 
Relying on some known results from group theory, we have written a basis of the Hilbert space in the form of semi-standard Young tableaux and we have derived their explicit second-quantization expression in order to discuss the effect of a loss process on them. 
This allowed us to write a lower bound for the average number of particles remaining in the lattice which should be satisfied at any time during the lossy dynamics, highlighting the fact that in general the system does not become empty at late time. 

In the second part of the paper, we have classified the dark states of the system according to the $(p,q)$-irrep of $\mathrm{SU}(3)$ they belong to, and we made a link with the weight diagram of the corresponding irreducible representation. We have shown that the majority of the dark states have $p+2q$ fermions. However, we also discussed the presence of dark states having a number of fermions strictly larger than $p+2q$ in the case of three particles of three different spin components $A$, $B$, and $C$.

A promising direction for future research is the exploration of the lossy dynamics that occur before the system reaches a stationary state. Such a dynamics is expected to be strongly constrained by the large number of conserved quantities that we have identified. 
Thus, it seems reasonable to assume that the density matrix of the gas belongs to a time-dependent generalized Gibbs ensemble, in the weak dissipation regime and thermodynamic limit, like it was done in related problems~\cite{
Bouchoule2020, Rosso2021, Riggio2024,ulcakar2024,lumia2024,lehr2025}. This assumption could be used to develop mean-field dynamical equations for observables.   

Finally, we expect that simulating the dynamics in the symmetry-resolved basis of semi-standard Young tableaux states, rather than in the usual Fock-state basis, could drastically reduce the computational cost of a stochastic quantum trajectory algorithm. In particular, this approach would naturally decouple the dynamics across the different $(p,q)$-sectors, allowing them to be computed in parallel.

%
%
\begin{acknowledgments}

We thank Gianni~Aupetit-Diallo, Yvan Castin, John Huckans, Bruno Laburthe-Tolra, Pierre Nataf, Benjamin Pasquiou, Martin Robert de Saint Vincent, and Aleksandra A. Ziolkowska for enlightening discussions.
This work was carried out in the framework of the joint Ph.D.\ program between the CNRS and the University of Tokyo. This work is supported by the ANR project LOQUST ANR-23-CE47-0006-02.
This work is part of HQI (www.hqi.fr) initiative and is supported by France 2030 under the French National Research Agency grant number ANR-22-PNCQ-0002.
H.K. was supported by JSPS KAKENHI Grants No. JP23K25783, No. JP23K25790, and MEXT KAKENHI Grant-in-Aid for Transformative Research Areas A “Extreme Universe” (KAKENHI Grant No. JP21H05191).

\end{acknowledgments}
%
%
%
%

\appendix
%
%
%
\section{Strong symmetries of the dissipative dynamics} \label{Ap:StrongSym}

In this Appendix, we prove that the operators $\Lambda_l$ are strong symmetries of the Lindbladian dynamics, described in Sec.~\ref{sec:model}. 

First, we note that any operator $\Lambda_l$ is a linear combination of the ladder operators $F_{\alpha \beta}$ defined in Eq.~\eqref{eq:Fab}. From Eqs.~\eqref{eq:Hhub2} and \eqref{eq:EF}, we know that $[H_{\rm Hub}, F_{\alpha \beta}] =0$. Therefore, we have
\begin{equation}
    [H_{\rm Hub}, \Lambda_l] = 0, \quad \forall l \in \{ 1,\ldots, 8 \}.
\end{equation}
On the other hand, since 
\begin{equation}
    [L_j, n_{i \sigma}] = \delta_{ij} L_j \quad \forall i,j \in \{ 1, \ldots,L \}, \quad \forall \sigma \in \{ A, B ,C\},
\end{equation}
we obtain directly that
\begin{equation}
    [L_j, \Lambda_3] = [L_j, \Lambda_8] = 0 \quad \forall j \in \{ 1, \ldots,L \}.
\end{equation}
 Additionally, we have
\begin{equation}
    [L_j,F_{\alpha \beta}] =0 \quad \text{for } \alpha \neq \beta.
\end{equation}
We note that any of the operators $\Lambda_{l^\prime}$ with  $l^\prime \in \{ 1, 2,4,5,6,7 \}$ is a linear combination of the operators $F_{\alpha \beta}$ with $\alpha \neq \beta$. Thus, $[L_j,\Lambda_{l^\prime}] =0$.

\section{Example of ABCD tableau and other quantum numbers} \label{Ap:ex_ABCD}

\begin{figure}[H]
\includegraphics[width=\linewidth]{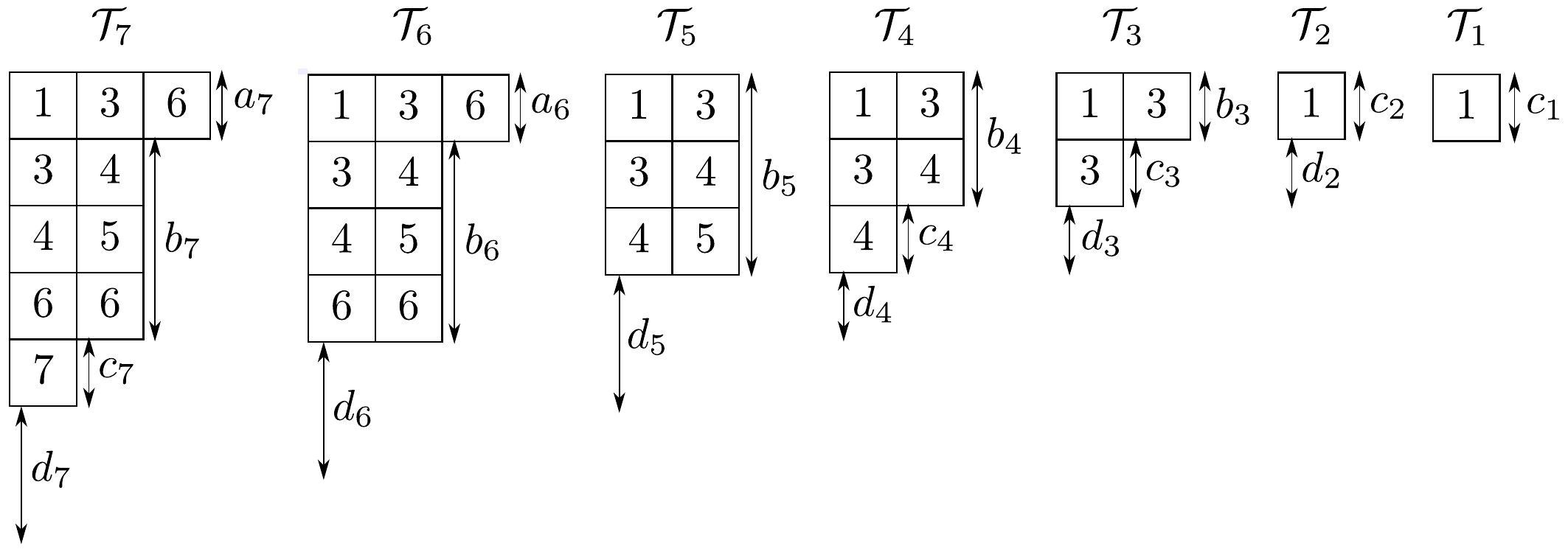}
\caption{Iterative procedure to find the quantum numbers $\{ a_k,b_k,c_k,d_k\}, \forall k \in \{1, \ldots, L \}$ for the orbital ssYT $\mathcal{T}_7$. In this example, we have $L=7$ lattice sites.} \label{fig:example_ABCD}
\end{figure}

In this Appendix, we illustrate how to construct the ABCD tableau introduced in Sec.~\ref{sec:secondQ} for a given orbital ssYT through an example. In Fig.~\ref{fig:example_ABCD}, we show the procedure for finding $\{ a_k,b_k,c_k,d_k \}, \forall k \in \{1, \ldots, L \}$ by iteratively removing some boxes. In Table~\ref{tab:example_ABCD}, we give the corresponding ABCD tableau and the quantum numbers $n_k$, $p_k$, and $q_k$.

\begin{table}[h!]
\begingroup
\renewcommand{\arraystretch}{2}
\begin{tabular}{c|c|c|c|c|c|c|c}
    \ $k$ \ & \ \shortstack{$a_k$} \ & \shortstack{$b_k = Q_k$} &  \shortstack{$c_k=P_k$} &  \ \shortstack{$d_k$} \ & \ \shortstack{$n_k$}  \ & \ \shortstack{$p_k$} \ & \ \shortstack{$q_k$} \ \\
   \hline \hline
 7  & 1 & 3 & 1 & 2 & 1 & 1 & 0 \\
 \hline
 6 & 1 & 3 & 0 & 2 & 3 & 0 & 0 \\
  \hline
 5 & 0 & 3 & 0 & 2 & 1 & 1 & 0 \\
  \hline
 4 & 0 & 2 & 1 & 1 & 2 & 0 & 1 \\
  \hline
 3 & 0 & 1 & 1 & 1 & 2 & 0 & 1 \\
  \hline
 2 & 0 & 0 & 1 & 1 & 0 & 0 & 0 \\
  \hline
 1 & 0 & 0 & 1 & 0 & 1 & 1 & 0 \\
\end{tabular}
\endgroup
\caption{All the quantum numbers characterizing the orbital ssYT $\mathcal{T}_7$ of Fig.~\ref{fig:example_ABCD}.}
\label{tab:example_ABCD}
\end{table}

\section{Examples for the second quantization writing of ssYT states} \label{ap:exSQ}

 In the Appendix, we illustrate how to explicitly apply Eq.~\eqref{eq:SDwriting} to obtain the second quantization writing of two specific states in Fig.~\ref{fig:example_state}(c), from their ssYT forms. Then, we provide the second quantization writing of the state in Fig.~\ref{fig:example_state}(b).

\underline{Example 1}: We aim to show that
\begin{equation*}
\begin{split}
 \frac{1}{\sqrt{3}} & \left( c_{2 A}^\dag c_{1 B}^\dag c_{1 C}^\dag +  c_{1 A}^\dag c_{2 B}^\dag c_{1 C}^\dag +  c_{1 A}^\dag c_{1 B}^\dag c_{2 C}^\dag \right) \ket{\rm vac} \\ =  \ &
 \begin{ytableau}
  1 & 1 & 2
 \end{ytableau} \ \begin{ytableau}
 A \\ B \\ C
 \end{ytableau} \ .
\end{split}
\end{equation*}
The quantum numbers associated to this state are $I_L=\Lambda_L^3 = \Lambda_L^8=0$ (deduced from the spin ssYT), and
\begin{table}[h!]
\begingroup
\renewcommand{\arraystretch}{1.13}
\begin{tabular}{c|c|c|c|c|c}
 \ $k$ \ & \ $P_k$ \ & \ $Q_k$ \ & \ $n_k$ \ & \ $p_k$ \ & \ $q_k$ \ \\
 \hline
 $2$ & $0$ & $0$ & $1$ & $1$ & $0$\\
  \hline
 $1$ & $0$ & $1$ & $2$ & $0$ & $1$
\end{tabular}
\endgroup
\end{table}

\noindent (deduced form the orbital ssYT). Applying Eq.~\eqref{eq:SDwriting} leads to
\begin{widetext}
\begin{equation*}
\begin{split}
& \begin{ytableau}
  1 & 1 & 2
 \end{ytableau} \ \begin{ytableau}
 A \\ B \\ C
 \end{ytableau}    =  \\ &  \underbrace{\bra{ 0 \ 0 \ 0 \ 0 \ 0 \ , 0 \ 1 \ 0 \ 0 \ \frac{1}{\sqrt{3}}} \ket{ \ 0 \ 1 \ 0 \ 0 \ \frac{1}{\sqrt{3}}}}_{=1} c_{1 A}^\dag c_{1 B}^\dag \underbrace{\bra{ 0 \ 1 \ 0 \ 0 \ \frac{1}{\sqrt{3}}  \ , 1 \ 0 \ 0 \ 0 \ \frac{-1}{\sqrt{3}}} \ket{ \ 0 \ 0 \ 0 \ 0 \ 0 }}_{=1/\sqrt{3}}   c_{2 C}^\dag  \ket{\rm vac}
 \\ + & \underbrace{\bra{ 0 \ 0 \ 0 \ 0 \ 0 \ , 0 \ 1 \ \frac{1}{2} \ \frac{1}{2} \ \frac{-1}{2 \sqrt{3}}} \ket{ \ 0 \ 1 \ \frac{1}{2} \ \frac{1}{2} \ \frac{-1}{2 \sqrt{3}}} }_{=1} c_{1 A}^\dag c_{1 C}^\dag \underbrace{\bra{ 0 \ 1 \ \frac{1}{2} \ \frac{1}{2} \ \frac{-1}{2 \sqrt{3}}  \ , 1 \ 0 \ \frac{1}{2} \ \frac{-1}{2} \ \frac{1}{2 \sqrt{3}}} \ket{ \ 0 \ 0 \ 0 \ 0 \ 0 }}_{=-1/\sqrt{3}} c_{2 B}^\dag   \ket{\rm vac}
 \\  + & \underbrace{\bra{ 0 \ 0 \ 0 \ 0 \ 0 \ , 0 \ 1 \ \frac{1}{2} \ \frac{-1}{2} \ \frac{-1}{2 \sqrt{3}}}\ket{\ 0 \ 1 \ \frac{1}{2} \ \frac{-1}{2} \ \frac{-1}{2 \sqrt{3}}} }_{=1}    c_{1 B}^\dag c_{1 C}^\dag \underbrace{\bra{ 0 \ 1 \ \frac{1}{2} \ \frac{-1}{2} \ \frac{-1}{2 \sqrt{3}} \ , 1 \ 0 \ \frac{1}{2} \ \frac{1}{2} \ \frac{1}{2 \sqrt{3}}} \ket{ \ 0 \ 0 \ 0 \ 0 \ 0 }}_{=1/\sqrt{3}}  c_{2 A}^\dag  \ket{\rm vac}.
\end{split}
\end{equation*}
We recall that some useful quantum numbers are given in Fig.~\ref{fig:state_cluster}. The $\mathrm{SU}(3)$ Clebsch–Gordan coefficients can either be found in standard tables \cite{McNAMEE, Kaeding} or computed using efficient algorithms \cite{Alex}.
\end{widetext}
\underline{Example 2}: We aim to show that
\begin{equation*}
 \frac{1}{\sqrt{2}}  \left( c_{1 A}^\dag c_{2 B}^\dag c_{2 C}^\dag -  c_{2 A}^\dag c_{1 B}^\dag c_{2 C}^\dag \right) \ket{\rm vac} =  \
 \begin{ytableau}
  1 & 2\\ 2
 \end{ytableau} \ \begin{ytableau}
 A & B \\ C
 \end{ytableau} \ .
\end{equation*}
The quantum numbers associated to this state are $I_L = 1$, $\Lambda_L^3 = \Lambda_L^8=0$ (deduced from the spin ssYT), and
\begin{table}[h!]
\begingroup
\renewcommand{\arraystretch}{1.13}
\begin{tabular}{c|c|c|c|c|c}
 \ $k$ \ & \ $P_k$ \ & \ $Q_k$ \ & \ $n_k$ \ & \ $p_k$ \ & \ $q_k$ \ \\
 \hline
 $2$ & $1$ & $1$ & $2$ & $0$ & $1$\\
  \hline
 $1$ & $1$ & $0$ & $1$ & $1$ & $0$
\end{tabular}
\endgroup
\end{table}

\noindent (deduced form the orbital ssYT). Applying Eq.~\eqref{eq:SDwriting} leads to
\begin{widetext}
\begin{equation*}
\begin{split}
& \begin{ytableau}
  1 & 2\\  2
 \end{ytableau} \ \begin{ytableau}
 A & B \\ C
 \end{ytableau}    =  \\ &  \underbrace{\bra{ 0 \ 0 \ 0 \ 0 \ 0 \ , 1 \ 0 \ \frac{1}{2} \ \frac{1}{2} \ \frac{1}{2 \sqrt{3}}} \ket{ \ 1 \ 0 \ \frac{1}{2} \ \frac{1}{2} \ \frac{1}{2\sqrt{3}}}}_{=1} c_{1 A}^\dag \underbrace{\bra{ \ 1 \ 0 \ \frac{1}{2} \ \frac{1}{2} \ \frac{1}{2\sqrt{3}} \ , 0 \ 1 \ \frac{1}{2} \ \frac{-1}{2} \ \frac{-1}{2\sqrt{3}}} \ket{ \ 1 \ 1 \ 1 \ 0 \ 0 }}_{=1/\sqrt{2}} c_{2 B}^\dag  c_{2 C}^\dag  \ket{\rm vac}
 \\ + &  \underbrace{\bra{ 0 \ 0 \ 0 \ 0 \ 0 \ , 1 \ 0 \ \frac{1}{2} \ \frac{-1}{2} \ \frac{1}{2 \sqrt{3}}} \ket{ \ 1 \ 0 \ \frac{1}{2} \ \frac{-1}{2} \ \frac{1}{2\sqrt{3}}}}_{=1} c_{1 B}^\dag \underbrace{\bra{ \ 1 \ 0 \ \frac{1}{2} \ \frac{-1}{2} \ \frac{1}{2\sqrt{3}} \ , 0 \ 1 \ \frac{1}{2} \ \frac{1}{2} \ \frac{-1}{2\sqrt{3}}} \ket{ \ 1 \ 1 \ 1 \ 0 \ 0 }}_{=1/\sqrt{2}} c_{2 A}^\dag  c_{2 C}^\dag  \ket{\rm vac}
  \\ + &  \underbrace{\bra{ 0 \ 0 \ 0 \ 0 \ 0 \ , 1 \ 0 \ 0 \ 0 \ \frac{-1}{\sqrt{3}}} \ket{ \ 1 \ 0 \ 0 \ 0 \ \frac{-1}{\sqrt{3}}}}_{=1} c_{1 C}^\dag \underbrace{\bra{ \ 1 \ 0 \ 0 \ 0 \ \frac{-1}{\sqrt{3}} \ , 0 \ 1 \ 0 \ 0 \ \frac{1}{\sqrt{3}}} \ket{ \ 1 \ 1 \ 1 \ 0 \ 0 }}_{=0} c_{2 A}^\dag  c_{2 B}^\dag  \ket{\rm vac}.
\end{split}
\end{equation*}
\end{widetext}
\underline{Example 3}: We now provide the second quantization writing of the state in Fig.~\ref{fig:example_state}(b). The associated quantum numbers are $I_L=\frac{1}{2}$, $\Lambda_L^3=-\frac{1}{2}$, $\Lambda_L^8 = \frac{1}{2 \sqrt{3}}$, and
\begin{table}[h!]
\begingroup
\renewcommand{\arraystretch}{1.13}
\begin{tabular}{c|c|c|c|c|c}
 \ $k$ \ & \ $P_k$ \ & \ $Q_k$ \ & \ $n_k$ \ & \ $p_k$ \ & \ $q_k$ \ \\
 \hline
 $4$ & $2$ & $1$ & $1$ & $1$ & $0$\\
  \hline
$3$ & $1$ & $1$ & $2$ & $0$ & $1$\\
  \hline
$2$ & $1$ & $0$ & $3$ & $0$ & $0$\\
  \hline
 $1$ & $1$ & $0$ & $1$ & $1$ & $0$
\end{tabular}
\endgroup
\end{table}

\noindent We have
\begin{equation*}
 \begin{split}
 \begin{ytableau}
  1 & 2 &2\\  2&3 \\ 3 \\ 4
 \end{ytableau} & \ \begin{ytableau}
 A & A & B  & C \\ B & B \\C
 \end{ytableau} =
 \\ &  \sqrt{\frac{3}{8}}  c_{1 C}^\dag c_{2 A}^\dag c_{2 B}^\dag c_{2 C}^\dag c_{3 A}^\dag c_{3 B}^\dag c_{4 B}^\dag \ket{\rm vac} \\
 + &  \sqrt{\frac{3}{8}}  c_{1 B}^\dag c_{2 A}^\dag c_{2 B}^\dag c_{2 C}^\dag c_{3 A}^\dag c_{3 B}^\dag c_{4 C}^\dag \ket{\rm vac} \\
 + &  \frac{1}{\sqrt{6}} c_{1 B}^\dag c_{2 A}^\dag c_{2 B}^\dag c_{2 C}^\dag c_{3 A}^\dag c_{3 C}^\dag c_{4 B}^\dag \ket{\rm vac}
 \\ - &  \frac{1}{2 \sqrt{6}} c_{1 A}^\dag c_{2 A}^\dag c_{2 B}^\dag c_{2 C}^\dag c_{3 B}^\dag c_{3 C}^\dag c_{4 B}^\dag \ket{\rm vac} \\
 -  &  \frac{1}{2 \sqrt{6}} c_{1 B}^\dag c_{2 A}^\dag c_{2 B}^\dag c_{2 C}^\dag c_{3 B}^\dag c_{3 C}^\dag c_{4 A}^\dag \ket{\rm vac}.
\end{split}
 \end{equation*}

\section{Proof of Eq.~\eqref{eq:LowerBound_SU3}} \label{Ap:Lower-Bound}
In this Appendix, we prove Eq.~\eqref{eq:LowerBound_SU3}. We recall that $\{ \ket{\varphi_i} \}$ is an orthonormal basis satisfying
    \begin{equation}
    \begin{split}
        &C_2 \ket{\varphi_i} = c_2(p_i,q_i) \ket{\varphi_i}, \quad C_3 \ket{\varphi_i} = c_3(p_i,q_i) \ket{\varphi_i},\\
        &\Nf \ket{\varphi_i} = \left( p_i + 2 q_i + 3 r_i \right) \ket{\varphi_i} , \ p_i,q_i,r_i \in \{0,1,\ldots\}        
    \end{split} 
    \end{equation}
First, we expand the density matrix of the system on this basis:
\begin{equation}
    \rho(t) = \sum_{ij} \rho_{ij}(t) \ket{\varphi_i} \bra{\varphi_j}, \quad \rho_{ij}(t) \in \mathbb{C}.
\end{equation}
The mean number of particles is 
\begin{align*}
    \Tr[\Nf \rho(t)] &= \sum_k \bra{\varphi_k} \Nf \rho(t) \ket{\varphi_k} \\
    &=  \sum_{ij}  \rho_{ij}(t) \sum_k \bra{\varphi_k} \Nf \ket{\varphi_i} \bra{\varphi_j} \ket{\varphi_k} \\
    &=  \sum_{ij}  \rho_{ij}(t)  \bra{\varphi_j} \Nf \ket{\varphi_i} \\
    &=  \sum_{ij}  \rho_{ij}(t) \left( p_i + 2 q_i +3 r_i \right) \bra{\varphi_j} \ket{\varphi_i} \\
     &=  \sum_{i}  \rho_{ii}(t) \left( p_i + 2 q_i +3 r_i \right) \geq \sum_{i}  \rho_{ii}(t) \left( p_i + 2 q_i \right)    \\
\end{align*}
The right-hand side of the inequality can be rewritten as
\begin{equation}
    \sum_{i}  \rho_{ii}(t) \left( p_i + 2 q_i \right) =  \sum_{p,q} \left( p + 2 q \right)  \sum_{\substack{i \text{ such that} \\ p_i=p \ q_i=q }}  \rho_{ii}(t).
\end{equation}
We remark that
\begin{equation}
      \sum_{\substack{i \text{ such that} \\ p_i=p \ q_i=q }}  \rho_{ii}(t)
\end{equation}
is the probability of being in a sector with given $p$ and~$q$. The latter quantity is time-independent since both $C_2$ and $C_3$ are strong symmetries. Thus, 
\begin{equation}
     \Tr[\Nf \rho(t)] \geq  \sum_{p,q} \left( p + 2 q \right) \sum_{\substack{i \text{ such that} \\ p_i=p \ q_i=q }}  \rho_{ii}(0)
\end{equation}
The latter equation is equivalent to Eq.~\eqref{eq:LowerBound_SU3}.
 
\section{Analytical derivation of the lower bounds for $\mathcal{N}$ in Fig.~\ref{fig:Few_part}} \label{ap:L3N3}
In this Appendix, we derive explicitly a lower bound of the total number of fermions remaining in the lattice for the two initial states considered in Fig.~\ref{fig:Few_part}:
\begin{subequations}
\begin{equation}
        \ket{\Psi_{0,1}} :=  c_{1A}^\dag c_{2B}^\dag c_{3C}^\dag \ket{\rm vac}, \label{Psi01}
\end{equation}
\begin{equation}
        \ket{\Psi_{0,2}} :=  \dfrac{1}{\sqrt{3}} \left( c_{1A}^\dag c_{2B}^\dag c_{3C}^\dag + c_{2A}^\dag c_{3B}^\dag c_{1C}^\dag + c_{3A}^\dag c_{1B}^\dag c_{2C}^\dag  \right) \ket{\rm vac}. \label{Psi02}
\end{equation}
\end{subequations}
For this, we will apply Eq.~\eqref{eq:LowerBound_SU3}.

First, we need to provide a basis of states $\ket{\varphi_i}$ diagonalizing both $C_2$ and $C_3$ for $N_A=N_B=N_C = 1$ on $L=3$ sites. We can conveniently write such basis states with their ssYT forms.
In the sector $(p=3,q=0,r=0)$, there exists only one state:
 \begin{subequations}
    \begin{align*}
    \frac{1}{\sqrt{6}}\Bigl( &c_{1A}^\dag  c_{2B}^\dag c_{3C}^\dag  - c_{1A}^\dag  c_{3B}^\dag c_{2C}^\dag + c_{2A}^\dag  c_{3B}^\dag c_{1C}^\dag\\ &- c_{2A}^\dag  c_{1B}^\dag c_{3C}^\dag  + c_{3A}^\dag  c_{1B}^\dag c_{2C}^\dag - c_{3A}^\dag  c_{2B}^\dag c_{1C}^\dag \Bigr) \ket{\rm vac}\\
    = & \ \begin{ytableau}
         1\\
         2\\
         3
    \end{ytableau} \ \begin{ytableau}
         A & B &  C
    \end{ytableau} \ := \ket{\varphi_1}.
    \end{align*}
\end{subequations}
In the sector $(p=1,q=1,r=0)$, four states have only single occupancies:
\begin{subequations}
 \begin{equation*}
     \begin{split}
  \frac{1}{2}  & \Bigl(   c_{3 A}^\dag   c_{2 B}^\dag c_{1 C}^\dag -  c_{2 A}^\dag c_{3 B}^\dag c_{1 C}^\dag + c_{3 A}^\dag c_{1 B}^\dag c_{2 C}^\dag   - c_{1 A}^\dag c_{3 B}^\dag c_{2 C}^\dag  \Bigr) \ket{\rm vac}  \\ & = \ \begin{ytableau}
         1 & 2\\
         3
    \end{ytableau}  \  \begin{ytableau}
         A & B \\
         C
    \end{ytableau} \ := \ket{\varphi_2},
 \end{split}
 \end{equation*}
  \begin{equation*}
     \begin{split}
  \frac{1}{2}  & \Bigl(   c_{3 A}^\dag   c_{2 B}^\dag c_{1 C}^\dag +  c_{2 A}^\dag c_{3 B}^\dag c_{1 C}^\dag - c_{3 A}^\dag c_{1 B}^\dag c_{2 C}^\dag - c_{1 A}^\dag c_{3 B}^\dag c_{2 C}^\dag  \Bigr) \ket{\rm vac}  \\ & = \ \begin{ytableau}
         1 & 3\\
         2
    \end{ytableau}  \  \begin{ytableau}
         A & C \\
         B
    \end{ytableau} \ := \ket{\varphi_3},
 \end{split}
 \end{equation*}
   \begin{equation*}
     \begin{split}
  \frac{1}{\sqrt{12}}  & \Bigl(  2 c_{1 A}^\dag   c_{2 B}^\dag c_{3 C}^\dag +  c_{1 A}^\dag c_{3 B}^\dag c_{2 C}^\dag - c_{3 A}^\dag c_{1 B}^\dag c_{2 C}^\dag \\ &  - 2 c_{2 A}^\dag c_{1 B}^\dag c_{3 C}^\dag - c_{2 A}^\dag c_{3 B}^\dag c_{1 C}^\dag + c_{3 A}^\dag c_{2 B}^\dag c_{1 C}^\dag \Bigr) \ket{\rm vac}  \\ & = \ \begin{ytableau}
         1 & 3\\
         2
    \end{ytableau}  \  \begin{ytableau}
         A & B \\
         C
    \end{ytableau} \ := \ket{\varphi_4},
 \end{split}
 \end{equation*}
    \begin{equation*}
     \begin{split}
  \frac{1}{\sqrt{12}}  & \Bigl(  2 c_{1 A}^\dag   c_{2 B}^\dag c_{3 C}^\dag -  c_{1 A}^\dag c_{3 B}^\dag c_{2 C}^\dag - c_{3 A}^\dag c_{1 B}^\dag c_{2 C}^\dag \\ &  + 2 c_{2 A}^\dag c_{1 B}^\dag c_{3 C}^\dag - c_{2 A}^\dag c_{3 B}^\dag c_{1 C}^\dag - c_{3 A}^\dag c_{2 B}^\dag c_{1 C}^\dag \Bigr) \ket{\rm vac}  \\ & = \ \begin{ytableau}
         1 & 2\\
         3
    \end{ytableau}  \  \begin{ytableau}
         A & C \\
         B
    \end{ytableau} \ := \ket{\varphi_5},
 \end{split}
 \end{equation*}
\end{subequations}
and twelve states have a double occupancy:
\begin{subequations}
 \begin{equation*}
  \frac{1}{\sqrt{2}}\left(c_{i A}^\dag c_{j B}^\dag c_{j C}^\dag - c_{j A}^\dag c_{i B}^\dag c_{j C}^\dag \right) \ket{\rm vac} = \ \begin{ytableau}
         j & j\\
         i
    \end{ytableau}  \  \begin{ytableau}
         A & B \\
         C
    \end{ytableau} \ ,
 \end{equation*}
 \begin{equation*}
  \frac{1}{\sqrt{2}}\left(c_{j A}^\dag c_{i B}^\dag c_{i C}^\dag - c_{i A}^\dag c_{j B}^\dag c_{i C}^\dag \right) \ket{\rm vac} = \ \begin{ytableau}
         j & i\\
         i
    \end{ytableau}  \  \begin{ytableau}
         A & B \\
         C
    \end{ytableau} \ ,
 \end{equation*}
 \begin{equation*}
 \begin{split}
  \frac{1}{\sqrt{6}} & \left( 2 c_{j A}^\dag c_{j B}^\dag c_{i C}^\dag - c_{i A}^\dag c_{j B}^\dag c_{j C}^\dag -  c_{j A}^\dag c_{i B}^\dag c_{j C}^\dag \right) \ket{\rm vac} \\ &= \ \begin{ytableau}
         j & j\\
         i
    \end{ytableau}  \  \begin{ytableau}
         A & C \\
         B
    \end{ytableau} \ ,
 \end{split}
 \end{equation*}
  \begin{equation*}
 \begin{split}
  \frac{1}{\sqrt{6}} & \left( 2 c_{i A}^\dag c_{i B}^\dag c_{j C}^\dag - c_{j A}^\dag c_{i B}^\dag c_{i C}^\dag -  c_{i A}^\dag c_{j B}^\dag c_{i C}^\dag \right) \ket{\rm vac} \\ &= \ \begin{ytableau}
         j & i\\
         i
    \end{ytableau}  \  \begin{ytableau}
         A & C \\
         B
    \end{ytableau} \ ,
 \end{split}
 \end{equation*}
\end{subequations}
with $i ,j \in \{1,2,3\}$ and $j<i$.
In the sector $(p=0,q=0,r=1)$, there are ten states. For three of them, all the fermions are on the same lattice site:
\begin{subequations}
    \begin{align*}
c_{jA}^\dag  c_{jB}^\dag c_{jC}^\dag \ket{\rm vac}   = \begin{ytableau}
         j & j &  j
    \end{ytableau} \ \begin{ytableau}
         A\\
         B\\
         C
    \end{ytableau} \mbox{ with } j \in \{1,2,3\}.
    \end{align*}
\end{subequations}
For six of them, exactly two fermions are on the same site:
\begin{subequations}
    \begin{align*}
\frac{1}{\sqrt{3}} & \left(  c_{iA}^\dag  c_{jB}^\dag c_{jC}^\dag + c_{jA}^\dag  c_{iB}^\dag c_{jC}^\dag  + c_{jA}^\dag  c_{jB}^\dag c_{iC}^\dag \right) \ket{\rm vac}
        \\ &= \begin{cases} \  \begin{ytableau}
         i & j &  j
    \end{ytableau} \ \begin{ytableau}
         A\\
         B\\
         C
    \end{ytableau} \quad \text{ if } i < j \\ \\ \ \begin{ytableau}
         j & j &  i
    \end{ytableau} \ \begin{ytableau}
         A\\
         B\\
         C
    \end{ytableau} \quad \text{ if } j < i \\ \end{cases} \mbox{ with } i,j \in \{1,2,3\}.
    \end{align*}
\end{subequations}
Finally for one of them, all the fermions are on different sites:

 \begin{subequations}
    \begin{align*}
  \frac{1}{\sqrt{6}} \Bigl( &c_{1A}^\dag  c_{2B}^\dag c_{3C}^\dag  + c_{1A}^\dag  c_{3B}^\dag c_{2C}^\dag + c_{2A}^\dag  c_{3B}^\dag c_{1C}^\dag \\ &+ c_{2A}^\dag  c_{1B}^\dag c_{3C}^\dag  + c_{3A}^\dag  c_{1B}^\dag c_{2C}^\dag + c_{3A}^\dag  c_{2B}^\dag c_{1C}^\dag \Bigr) \ket{\rm vac}\\
    = \ &  \begin{ytableau}
         1 & 2 & 3
    \end{ytableau}  \  \begin{ytableau}
         A \\
         B\\
         C
    \end{ytableau} \ := \ket{\varphi_6} .
    \end{align*}
\end{subequations}
For the two examples that we chose, the initial state $\rho(0) = \ket{\Psi_0} \bra{\Psi_0}$ contains only single lattice site occupancies, thus the only basis states having an overlap with it are $\ket{\varphi_i}$ where $i=1,\ldots,6$. By applying Eq.~\eqref{eq:LowerBound_SU3}, we obtain
\begin{equation*}
\begin{split}
 \forall t, \quad \mathcal{N}(t) & \geq \sum_{i=1}^6 (p_i + 2 q_i) \bra{\varphi_i} \rho(0)  \ket{\varphi_i} \\ & \geq 3 \times \sum_{i=1}^5 \bra{\varphi_i} \rho(0)  \ket{\varphi_i} + 0 \times  \bra{\varphi_6} \rho(0)  \ket{\varphi_6}
\end{split}
\end{equation*}
For $\ket{\Psi_0} = \ket{\Psi_{0,1}}$, defined in Eq.~\eqref{Psi01}, we obtain $\mathcal{N}(t) \geq 3 \times \left( \frac{1}{6} + 0 + 0 + \frac{4}{12} + \frac{4}{12} \right) = \frac{5}{2}$. On the other hand, for $\ket{\Psi_0} = \ket{\Psi_{0,2}}$, defined in Eq.~\eqref{Psi02}, we find $\mathcal{N}(t) \geq 3 \times \left( \frac{1}{2} + 0 + 0 + 0 + 0 \right) = \frac{3}{2}$.
\section{Expression of $p+2q$ as function of $c_2$ and $c_3$} \label{Ap:c1_c2_p_q}

In this Appendix, we show how Eq.~\eqref{eq:c1_c2_p_q} is obtained.
We first note the following: if $\tilde{N} = p+2q$ then
\begin{equation}
\begin{split}
         \frac{\tilde{N}^3}{9} - \tilde{N} c_2(p,q) + 2 c_3 (p,q) - 3 c_2(p,q)  + \tilde{N}^2 + 2 \tilde{N} =0, \label{eq:Cubic}
\end{split}
\end{equation}
where $c_2(p,q)$ and $c_3(p,q)$ are defined in Eqs.~\eqref{eq:c1} and~\eqref{eq:c2}. 
The discriminant of the latter third order polynomial equation is $\Delta = (1+p)^2 (1+q)^2 (2+p+q)^2 /9>0$. Thus, the three solutions, denoted $\tilde{N}_0$, $\tilde{N}_1$ and $\tilde{N}_2$, are distinct and real.
The corresponding depressed cubic form is 
\begin{equation}
    t^3 - 9 \left( 1+c_2\right) t+ 18 c_3 =0, \quad t = \tilde{N} + 3. \label{eq:depressedCubic}
\end{equation}
According to Galois theory, the three roots of Eq.~\eqref{eq:depressedCubic} are
\begin{equation}
\begin{split}
        t_k = 2 & \sqrt{3} \sqrt{1+c_2} \times \\ & \cos \left[  \frac{1}{3} \arccos \left( - \sqrt{3} \frac{c_3}{\left( 1+c_2 \right) ^{3/2}} \right) - \frac{2 \pi}{3} k \right],
\end{split}
\end{equation}
with $k=0,1,2$. Thus, the roots of the original cubic equation~\eqref{eq:Cubic} are $\tilde{N}_k = t_k -3$.
For $p\geq0$ and $q\geq0$, we verify that $p+2q = \tilde{N}_0$.

\section{Proof that a right-eigenvector of $H_{\rm eff}$ with real eigenvalue is dark} \label{Ap:DSProof}

In this Appendix, we prove that a right-eigenvector of $H_{\rm eff} = H_{\rm Hub} - \frac{i}{2} \sum_j L_j^\dag L_j$ with real eigenvalue is a dark state. Let $\ket{\Psi}$ be such that $H_{\rm eff} \ket{\Psi} = E \ket{\Psi}$ with ${\rm Im}(E)=0$.
We have 
\begin{equation}
    \bra{\Psi} H_{\rm Hub} \ket{\Psi} - \frac{i}{2} \sum_j \bra{\Psi} L_j^\dag L_j \ket{\Psi} = \underbrace{E \braket{\Psi}}_{\in  \mathbb{R}}.
\end{equation}
Since $H_{\rm Hub}$ and $\sum_j L_j^\dag L_j$ are both Hermitians, $\bra{\Psi} H_{\rm Hub} \ket{\Psi}$ and $\sum_j \bra{\Psi} L_j^\dag L_j \ket{\Psi}$ are real. Thus,
\begin{equation}
    \sum_j \bra{\Psi} L_j^\dag L_j \ket{\Psi} =0.
\end{equation}
However, $\bra{\Psi} L_j^\dag L_j \ket{\Psi} \geq 0,\forall j$, therefore $\bra{\Psi} L_j^\dag L_j \ket{\Psi} = \lVert L_j \ket{\Psi} \rVert^2 = 0,\forall j$, with $\lVert . \rVert$ the vector norm. Finally, $L_j \ket{\Psi} =0,\forall j$ which also implies that $H_{\rm Hub} \ket{\Psi}=E \ket{\Psi}$. Therefore, $\ket{\Psi}$ is dark.

\section{Proof of the link between stationary states and dark states expressed in Eq.~\eqref{eq:rhoDS}} \label{App:Eq:rhoDS}

In this Appendix, we prove Eq.~\eqref{eq:rhoDS}.
We note that the Liouvillian superoperator $\mathcal{L}$, defined in Eq.~\eqref{eq:MElattice}, can be formally diagonalized~\cite{torres2014closed}. We denote $\rho_\lambda$ its eigenoperators and $\lambda \in \mathbb{C}$ the associated eigenvalues: $\mathcal{L}[\rho_\lambda] = \lambda \rho_\lambda$. 
The initial density matrix of the system can be expanded as $\rho_0 = \sum_\lambda c_\lambda \rho_\lambda$, with $c_\lambda \in \mathbb{C}$. When $t \rightarrow \infty$, the density matrix becomes $\rho_{\infty} = \sum_{\lambda, \rm Re{(\lambda)} =0 } c_\lambda e^{\lambda t} \rho_\lambda$. 
Remarkably, for purely lossy systems, it is possible to prove that any eigenvalue of $\mathcal{L}$ satisfies the inequality $\rm Re ( \lambda)\leq 0 , \forall \lambda$, and it is written as $\lambda = -i \left(E^\prime-E^* \right)$ with $E$ and $E^\prime$ two eigenvalues of $H_{\rm eff}$~\cite{Nakagawa2021}.

Given the form of $H_{\rm eff}$ in Eq.~\eqref{Eq:30}, it is always true that $\mathrm{Im}(E)\leq 0$.
Hence, $\mathrm{Re}(\lambda)=0$ is possible only when $E'$ and $E$ are both real, the condition that in the main text we have associated to dark states.
With few algebraic passages, we can show that if $\ket{\Psi_i}$ and $\ket{\Psi_j}$ are both dark states of energies $E_i$ and $E_j$, then
$\mathcal{L}[\ket{\Psi_j} \bra{\Psi_i}] = -i \left( E_j - E_i \right) \ket{\Psi_j} \bra{\Psi_i}$. Thus, the eigenoperators $\rho_\lambda$, such that  ${\rm{Re}}(\lambda) = 0$, appearing in the expansion of $\rho_{\infty}$, can be identified as the operators $\ket{\Psi_j} \bra{\Psi_i}$ with purely imaginary eigenvalues $ - i \left( E_j - E_i \right)$.

\section{Details on the numerical exact diagonalization of $H_{\rm eff}$ in each $(p,q,I,r=0)$-sector} \label{App:ED}

In this Appendix, we provide some details concerning the numerical method used to find the dark states containing $\Nf = p+2q$ fermions. First, we note that the effective non-Hermitian Hamiltonian $H_{\rm eff} = H_{\rm Hub} - i \frac{\gamma}{2} \sum_j L_j^\dag L_j$, the $\mathrm{SU}(2)$ and $\mathrm{SU}(3)$ Casimir operators $I^2$, $C_2$, $C_3$, and the spin-resolved particle number operators, $N_A$, $N_B$, $N_C$ mutually commute, so that they can be diagonalized in the same basis. We construct the matrix representation of $H_{\rm eff}$, $I^2$, $C_2$, and $C_3$ in the basis of Fock states $| F^{(n_A)} \rangle \otimes | F^{(n_B)} \rangle \otimes| F^{(n_C)} \rangle$ for fixed numbers of $A$, $B$, and $C$ fermions equal to $n_A$, $n_B$, and $n_C$, respectively. 
Then, we perform an exact diagonalization of $H_{\rm eff} + \epsilon_1 I^2 + \epsilon_2 C_2 + \epsilon_3 C_3$ for some arbitrary choice of $\epsilon_1,\epsilon_2, \epsilon_3\in \mathbb{R}$. 
We check that the eigenvectors of this operator simultaneously diagonalize both $H_{\rm eff}$ and the Casimirs $I^2, C_2$ and $C_3$.

The eigenstates of $H_{\rm eff}$, $H_{\rm eff}\ket{\psi}=\epsilon\ket{\psi}$, such that ${\rm Im}\epsilon=0$ are the targeted dark states. To find the eigenvalue $\epsilon$ we compute the expectation value of the effective Hamiltonian, $\bra{\psi}H_{\rm eff} \ket{\psi} = \epsilon$. We next identify the symmetry sector $(p,q,I,r=0)$: by using Eqs.~\eqref{eq:c1} and~\eqref{eq:c2} we obtain the pair $(p,q)$ from the numerical eigenvalues $c_2(p,q)=\bra{\psi}C_2 \ket{\psi}$ and $c_3(p,q)=\bra{\psi}C_3 \ket{\psi}$, while we get $I$ from $I(I+1)=\bra{\psi}I^2 \ket{\psi}$.

\section{Proof of Eqs.~\eqref{eq:etaEV}} \label{Ap:eta}

In this Appendix, we prove the Eqs.~\eqref{eq:etaEV}, that are satisfied by the $\eta$-pairing states of the form~\eqref{eq:eta}.
By lengthy computation, it is possible to obtain the commutation relations
\begin{subequations}
    \begin{equation}
\left[ \eta_{\alpha \beta}^\dag, \eta_{\alpha^\prime \beta^\prime}^\dag \right] = 0, \quad \left[ H_{\rm hop}, \eta_{\alpha \beta}^\dag \right] = 0, 
\end{equation}
\begin{equation}
\left[ H_{\rm int,2}, \eta_{\alpha \beta}^\dag \right] = U_2 \eta_{\alpha \beta}^\dag + 2 U_2 R_{\alpha \beta \gamma} , 
\end{equation}
\begin{equation}
\left[ H_{\rm int,3}, \eta_{\alpha \beta}^\dag \right] = U_3 R_{\alpha \beta \gamma} , \quad R_{\alpha \beta \gamma} = \sum_j (-1)^j c_{j \alpha}^\dag c_{j \beta}^\dag n_{j \gamma},
\end{equation} \label{eq:commut}
\end{subequations}
with $\gamma \neq \alpha, \ \gamma \neq \beta$.
We also note that 
\begin{equation}
    R_{\alpha \beta \gamma} \ket{\eta; N_{\rm AB}, N_{\rm AC},N_{\rm BC}} = 0, \label{eq:RABC}
\end{equation}
because the state $\ket{\eta; N_{\rm AB}, N_{\rm AC},N_{\rm BC}}$ does not contain any singly-occupied lattice site, and thus the operators $c_{j \alpha}^\dag c_{j \beta}^\dag n_{j \gamma}$ annihilate it. Using Eqs.~\eqref{eq:commut} and~\eqref{eq:RABC}, we obtain Eqs.~\eqref{eq:etaEV}.

\section{Non-dissipative subspace for three bosons} \label{Ap:3bodyDSboson}

In this Appendix, we show that the subspace generated by the states~\eqref{eq:DS_generators} is closed under Hamiltonian evolution. We also give an expression of the dimension of this subspace; 
we stress that we are not able to give an explicit expression for the dark states living in this subspace.

We denote 
\begin{equation}
H_{\rm hop}^B = -J \sum_j b_{j+1}^\dag b_j + \rm{h.c.}    
\end{equation}
the hopping part of $H_{\rm Hub}^B$ and 
\begin{equation}
 H_{\rm int}^B= \frac{U_2}{2} \sum_j n_j (n_j-1)   
\end{equation}
its interaction part. With straightforward calculations, one can obtain that
\begin{widetext}
\begin{subequations}
    \begin{equation}
    \frac{H_{\rm hop}^B}{-J}   \ket{\Psi_1^\delta} =  \ket{\Psi_2^\delta} - 2 e^{i\delta}  \ket{\Psi_1^\delta},
    \end{equation}
    \begin{equation}
        \forall \Delta \geq 2, \quad  \frac{H_{\rm hop}^B}{-J}   \ket{\Psi_\Delta^\delta} =  \ket{\Psi_{\Delta-1}^\delta} +  \ket{\Psi_{\Delta+1}^\delta} + \sqrt{2}  \ket{\Phi_{\Delta}^\delta} + \sqrt{2} e^{i\delta} \ket{\Phi_{\Delta+1}^\delta} \quad \text{with } \ket{\Phi_{2}^\delta} :=0, 
    \end{equation}
    \begin{equation}
    \begin{split}
    \forall \Delta \geq 3, \quad  &\frac{H_{\rm hop}^B}{-J}   \ket{\Phi_\Delta^\delta} =  \sqrt{2} \ket{\Psi_{\Delta}^\delta} + \sqrt{2} e^{i \delta} \ket{\Psi_{\Delta-1}^\delta} + \ket{\Phi_{\Delta+1}^\delta} + \ket{\Phi_{\Delta-1}^\delta} + \ket{\Gamma_{2 \ \Delta}^\delta} + e^{i \delta} \ket{\Gamma_{2 \ \Delta+1}^\delta} \\ & \qquad  \text{with } \ket{\Gamma_{2 \ 3}^\delta} := - e^{i \delta} \ket{\Phi_{3}^\delta}, \quad  \ket{\Gamma_{\Delta \ 2 \Delta}^\delta} = \ket{\Gamma_{\Delta \ L-2 \Delta}^\delta}  = \ket{\Gamma_{2 \Delta \ L-2 \Delta}^\delta} = 0,
    \end{split}
    \end{equation} 
    \begin{equation}
    \begin{split}
           \forall \Delta_1 \geq 2, \ \Delta_2 \geq \Delta_1+2, \quad  \frac{H_{\rm hop}^B}{-J}   \ket{\Gamma_{\Delta_1 \ \Delta_2}^\delta} =&  \ket{\Gamma_{\Delta_1 \ \Delta_2+1}^\delta} +  \ket{\Gamma_{\Delta_1 \ \Delta_2-1}^\delta} +  \ket{\Gamma_{\Delta_1+1 \ \Delta_2}^\delta} +  \ket{\Gamma_{\Delta_1-1 \ \Delta_2}^\delta} \\&+  e^{i \delta} \ket{\Gamma_{\Delta_1-1 \ \Delta_2-1}^\delta} + e^{i \delta}  \ket{\Gamma_{\Delta_1+1 \ \Delta_2+1}^\delta}.
    \end{split}
    \end{equation}
\end{subequations}
\end{widetext}
We also have
\begin{subequations}
 \begin{equation}    
    \forall \Delta \geq 1, \quad H_{\rm int}^B/U_2 \ket{\Psi_\Delta^\delta} =\ket{\Psi_\Delta^\delta},
\end{equation}
\begin{equation}
    \forall \Delta \geq 3, \quad H_{\rm int}^B/U_2 \ket{\Phi_\Delta^\delta} =0,
\end{equation}
\begin{equation}
    \forall \Delta_1 \geq 2, \forall \Delta_2 \geq \Delta_1+2, \quad H_{\rm int}^B/U_2 \ket{\Gamma_{\Delta_1 \ \Delta_2}^\delta} = 0.
\end{equation}
\end{subequations}
Since the system size $L$ is finite, there exists a maximal value for $\Delta$, $\Delta_1$ and $\Delta_2$ so that the considered states are linearly independent. 

\begin{figure}[h!]
\includegraphics[width=\linewidth]{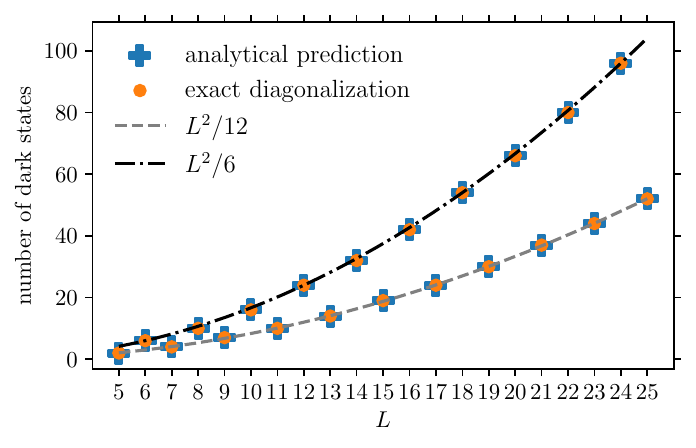}
\caption{Comparison between the number of dark states in the sector $(p=0,q=0,r=1)$ found by exact diagonalization of $H_{\rm Hub}^B$ and the analytical prediction of Eq.~\eqref{eq:NDS_boson} for $L \in \{5, \dots, 25\}$.} \label{fig:NDS_boson}
\end{figure}

We now assume that $L\geq 5$; the dark states for $L=3,4$ are given in Table~\ref{tab:DS}. For the states of the form $\ket{\Psi_\Delta^\delta}$, the maximal value of $\Delta$ is $\Delta_{\rm max} = L/2 -1$ if $L$ is even and $\Delta_{\rm max} =(L-1)/2$ if $L$ is odd. For the states of the form $\ket{\Phi_\Delta^\delta}$, we have $\Delta_{\rm{max}}= L/2$ if $L$ is even and   $\Delta_{\rm{max}} = (L-1)/2$ if $L$ is odd. Finally, for $\ket{\Gamma_{\Delta_1 \ \Delta_2}^\delta}$, the state can be characterized by three non-independent numbers $d_1$, $d_2$, and $d_3$ which are the distances between two neighboring particles. For instance, we have $d_1 = \Delta_1$, $d_2 = \Delta_2 -\Delta_1$, and $d_3 = L - 2 \Delta_1-\Delta_2$. The number of $\ket{\Gamma_{\Delta_1 \ \Delta_2}^\delta}$ states (for $\delta$ fixed) is equal to the number of possible triplets of integers $(d_1,d_2,d_3)$ such that $1<d_1 < d_2<d_3 $ and $d_1+d_2+d_3=L$, which is $f(L) = \sum_{d_1=2}^L \sum_{d_2=d_1+1}^L \sum_{d_3=d_2+1}^L \delta_{L,d_1+d_2+d_3}$. An analytical expression for $f(L)$ can be derived:
\begin{equation*}
\begin{split}
&\forall L <9, \ f(L)=0 ,\\ 
        &\forall L \geq 9, \
    f(L)=   \begin{cases}
        \frac{(L-6)^2}{12} &\text{ for } L= 0 \bmod 6, \\ \\
        \frac{(L-5)(L-7)}{12} &\text{ for } L= 1 \text{ or } 5 \bmod 6 ,\\ \\
        \frac{(L-4)(L-8)}{12} &\text{ for } L= 2 \text{ or } 4 \bmod 6 ,\\ \\
        \frac{39+L(L-12)}{12}  &\text{ for } L= 3 \bmod 6 .
    \end{cases}
\end{split}
\end{equation*}
Therefore, the dimension of the non-dissipative subspace is
\begin{equation}
    \begin{split}
        2 \left( \frac{L}{2} - 1 \right) + 2 \left( \frac{L}{2} - 2 \right) + 2 f(L) \text{ if } L \text{ is even}, \\
         \frac{L-1}{2} +  \frac{L-1}{2} -2 + f(L) \text{ if } L \text{ is odd}. \label{eq:NDS_boson}
    \end{split}
\end{equation}
In Fig.~\ref{fig:NDS_boson}, we show the comparison between the number of dark states in the sector $(p=0,q=0,r=1)$, obtained via exact diagonalization of $H_{\rm Hub}^B$, and the prediction of  Eq.~\eqref{eq:NDS_boson} for $L \in \{5, \dots, 25\}$. We obtain a perfect agreement.

%
%
%
\newpage
\bibliography{biblio}
\end{document}